\documentclass[11pt, letterpaper]{article}
\usepackage[utf8]{inputenc}
\usepackage[english]{babel}
\usepackage{amsmath}
\usepackage{mathtools}
\usepackage{amsfonts}
\usepackage{amssymb}
\usepackage[round,numbers,sort&compress]{natbib}
\usepackage{graphicx}
\usepackage[dvipsnames]{xcolor}
\usepackage{parskip}
\usepackage{tabu}
\usepackage{geometry} 
\usepackage{verbatim} 
\usepackage{helvet,times}
\usepackage{afterpage} 

\usepackage{multibib}
\newcites{supp}{Supporting References}

\usepackage{newfloat}
\DeclareFloatingEnvironment[name={Supporting Figure}]{suppfigure}
\DeclareFloatingEnvironment[name={Supporting Movie}]{suppmovie}
\DeclareFloatingEnvironment[name={Supporting Table}]{supptable}

\newcommand{\revise}[1]{#1}  
\newcommand{\hf}{\fontfamily{phv}\selectfont}{}  

\begin{document}

\title{\flushleft \hf {\bf \hf
 Pulsatile lipid vesicles under osmotic stress}\\ \bigskip
 {\normalsize Morgan Chabanon$^\text{a}$, James C.S. Ho$^\text{b}$, Bo Liedberg$^\text{b}$, Atul N. Parikh$^\text{b,c}$, Padmini Rangamani$^{\text{a}*}$} \\ \bigskip 
{\small $^\text{a}$ Department of Mechanical and Aerospace Engineering, University of California San Diego, La Jolla, CA, USA. \\
$^\text{b}$ Center for Biomimetic Sensor Science, School of Materials Science and Engineering, Nanyang Technological University, Singapore. \\
$^\text{c}$ Departments of Applied Science, Biomedical Engineering, and Chemical Engineering and Materials Science, University of California Davis, Davis, CA, USA.\\
\textit{$^*$ Corresponding author: prangamani@eng.ucsd.edu}}}

\author{ }
\date{}

\maketitle

{\hf ABSTRACT }
The response of lipid bilayers to osmotic stress is an important part of cellular function. Recent experimental studies showed that when cell-sized giant unilamellar vesicles (GUVs) are exposed to hypotonic media, they respond to the osmotic assault by undergoing a cyclical sequence of swelling and bursting events, coupled to the membrane's compositional degrees of freedom. Here, we establish a fundamental and quantitative understanding of the essential pulsatile behavior of GUVs under hypotonic conditions by advancing a comprehensive theoretical model of vesicle dynamics. The model quantitatively captures the experimentally measured swell-burst parameters for single-component GUVs, and reveals that thermal fluctuations enable rate-dependent pore nucleation, driving the dynamics of the swell-burst cycles. We further extract constitutional scaling relationships between the pulsatile dynamics and GUV properties over multiple time scales. Our findings provide a fundamental framework that has the potential to guide future investigations on the non-equilibrium dynamics of vesicles under osmotic stress.

\section*{\hf INTRODUCTION}

In their constant struggle with the environment, living cells of contemporary organisms employ a variety of highly sophisticated molecular mechanisms to deal with sudden changes in their surroundings. One often encountered environmental assault on cells is osmotic stress, where the amount of dissolved molecules in the extracellular environment drops suddenly \cite{christensen1987, hoffmann2009}. If left unchecked, this perturbation will result in a rapid flow of water into the cell through osmosis, causing it to swell, rupture, and die. To avoid this catastrophic outcome, even bacteria have evolved complex molecular machineries, such as mechanosensitive channel proteins, which allow them to release excess water from their interior \cite{berrier1996, blount1997, levina1999, wood1999}. This then raises an intriguing question of how might primitive cells, or cell-like artificial constructs, that lack the sophisticated protein machinery for osmosensing and osmoregulation, respond to such environmental insults and preserve their structural integrity.
 
Using rudimentary cell-sized giant unilamellar vesicles (GUVs) devoid of proteins and consisting of amphiphilic lipids and cholesterol as models for simple protocells, we showed previously that vesicular compartments respond to osmotic assault created by the exposure to hypotonic media by undergoing a cyclical sequence of swelling and poration \cite{oglecka2014}. In each cycle, osmotic influx of water through the semi-permeable boundary swells the vesicles and renders the bounding membrane tense, which in turn, opens a microscopic transient pore, releasing some of the internal solutes before resealing. This swell-burst process, \revise{depicted in Fig.~\ref{fig1}(A), }repeats multiple times producing a pulsating pattern in the size of the vesicle undergoing osmotic relaxation. From a dynamical point of view, this autonomous osmotic response results from an initial, far-from-equilibrium, thermodynamically unstable state generated by the sudden application of osmotic stress. The subsequent evolution of the system, characterized by the swell-burst sequences described above, occurs in the presence of a global constraint, namely constant membrane area, during a dissipation-dominated process \cite{peterlin2008, ho2016}.  

The study of osmotic response of lipid vesicles has a rich history in theoretical biophysics, beginning with the pioneering work by Koslov \& Markin \cite{koslov1984}, who provided some of the early theoretical foundations of osmotic swelling of lipid vesicles. 
In this work, they predicted that the response of a sub-micrometer sized vesicles to osmotic stress is likely pulsatile and due to the formation of successive transient pores (see Fig.~9, in \cite{koslov1984}, for a schematic for the volume change of the vesicle over time). They further approximated the characteristic quantities of swell-burst cycles (e.g. swelling time, critical volumes), based on the probability of the membrane overcoming the nucleation energy barrier to form a pore.
Independently, the dynamics of a single transient pore in a tense membrane were first theorized by Litster \cite{litster1975}, and later investigated theoretically and experimentally by Brochard-Wyart and coworkers \cite{sandre1999, brochard-wyart2000}. Idiart \& Levin \cite{idiart2004} combined the osmotic swelling theory and pore dynamics, and calculated the dynamics of a pulsatile behavior assuming a constant lytic tension. 
These modeling efforts made great strides in our understanding of some of the essential physics underlying vesicle responses to osmotic stress. 

Previously, we used these ideas to provide a qualitative interpretation of pulsatile behavior of GUVs (see schematics in \cite{oglecka2014} Fig.~7h,i). However a general framework that quantitatively describes the response of pulsatile vesicles to osmotic stress at all relevant time scales is still missing. The success of such a model must rely on (a) the integration of vesicle dynamics, pore dynamics with nucleation, and long-time solute concentration dynamics within a unified framework, and (b) the assessment of the model predictions with respect to experimental measurements, in order to establish the physical relevance of the essential parameters that govern the system dynamics. Here, we build on the findings and theories reported previously \cite{ litster1975, koslov1984, brochard-wyart2000, idiart2004, evans2003, ryham2011} to develop such a quantitative model for the dynamics of swell-burst cycles in giant lipid vesicles subject to osmotic stress. 

In analyzing the pulsatile dynamics of GUVs, a number of general questions naturally arise: (i) Is the observed condition for membrane poration deterministic or stochastic? (ii) Is poration controlled by a unique value of membrane tension (\textit{i.e.} lytic tension) introduced by the area-volume changes, which occur during osmotic influx, or does it involve coupling of the membrane response to thermal fluctuations? (iii) Does the critical lytic tension depend on the strain rate, and thus the strength of the osmotic gradient? Such questions arise beyond the present context of vesicle osmoregulation in other important scenarios where the coupling between the dissipation of osmotic energy and cellular compartmentalization has important biological ramifications \cite{rand2004, diz-munoz2013, stroka2014, porta2015}. 

Motivated by these considerations, we carried out a combined theoretical-experimental study integrating membrane elasticity, continuum transport, and statistical thermodynamics. We gathered quantitative experimental data to address the questions above, and developed a general model 
that recapitulates the essential qualitative features of the experimental observations, emphasizes the importance of dynamics, and places the heretofore neglected contribution of thermal fluctuations in driving osmotic response of stressed vesicular compartments.

\section*{\hf MATERIALS AND METHODS}

The detailed materials and methods used in this work are available in Supporting Materials and Methods in the Supporting Material. The experimental configuration is similar to that already described \cite{angelova1992, oglecka2014}. Briefly, we prepared GUVs consisting essentially of a single amphiphile, namely 1-palmitoyl-2-oleoyl-sn-1-glycero-3-phosphocholine (POPC), doped with a small concentration (1~mol$\%$) of a fluorescently labeled phospholipid (1,2-dipalmitoyl -sn-glycero-3-phosphoethanolamine-N-(lissamine rhodamine B sulfonyl)) or Rho-DPPE using standard electroformation technique \cite{angelova1992}. The GUVs thus obtained were typically between 7 and 20~$\mu$m in radius, encapsulated 200~mM sucrose, and were suspended in the isotonic glucose solution of identical osmolarity. 
Diluting the extra-vesicular dispersion medium with deionized water produces a hypotonic bath depleted in osmolytes, subjecting the GUVs to osmotic stress.  Shortly ($\sim$1~min delay) after subjecting the GUVs to the osmotic differential, GUVs were monitored using time-lapse epifluorescence microscopy at a rate of 1~image per 150~ms, and images were analyzed using a customized MATLAB code to extract the evolution of the GUV radii with time, with a precision of about 0.1~$\mu$m.

We developed a mathematical model predicting the pulsatile behavior of GUVs in hypotonic environment. Essentially, the model couples pore nucleation by thermal fluctuations, osmotic swelling, and solute transport. These aspects are represented by Eqs.~\ref{eq:sde_r}, \ref{eq:ode_R} and \ref{eq:ode_c} respectively, and discussed below. Details regarding the theory and its numerical implementation are reported in Model Development and Simulations in the Supporting Materials.

\section*{\hf RESULTS}

\subsection*{\hf Homogeneous GUVs display swell-burst cycles in hypotonic conditions}

\begin{figure*}[tb] 
\centering
\includegraphics[scale=0.55]{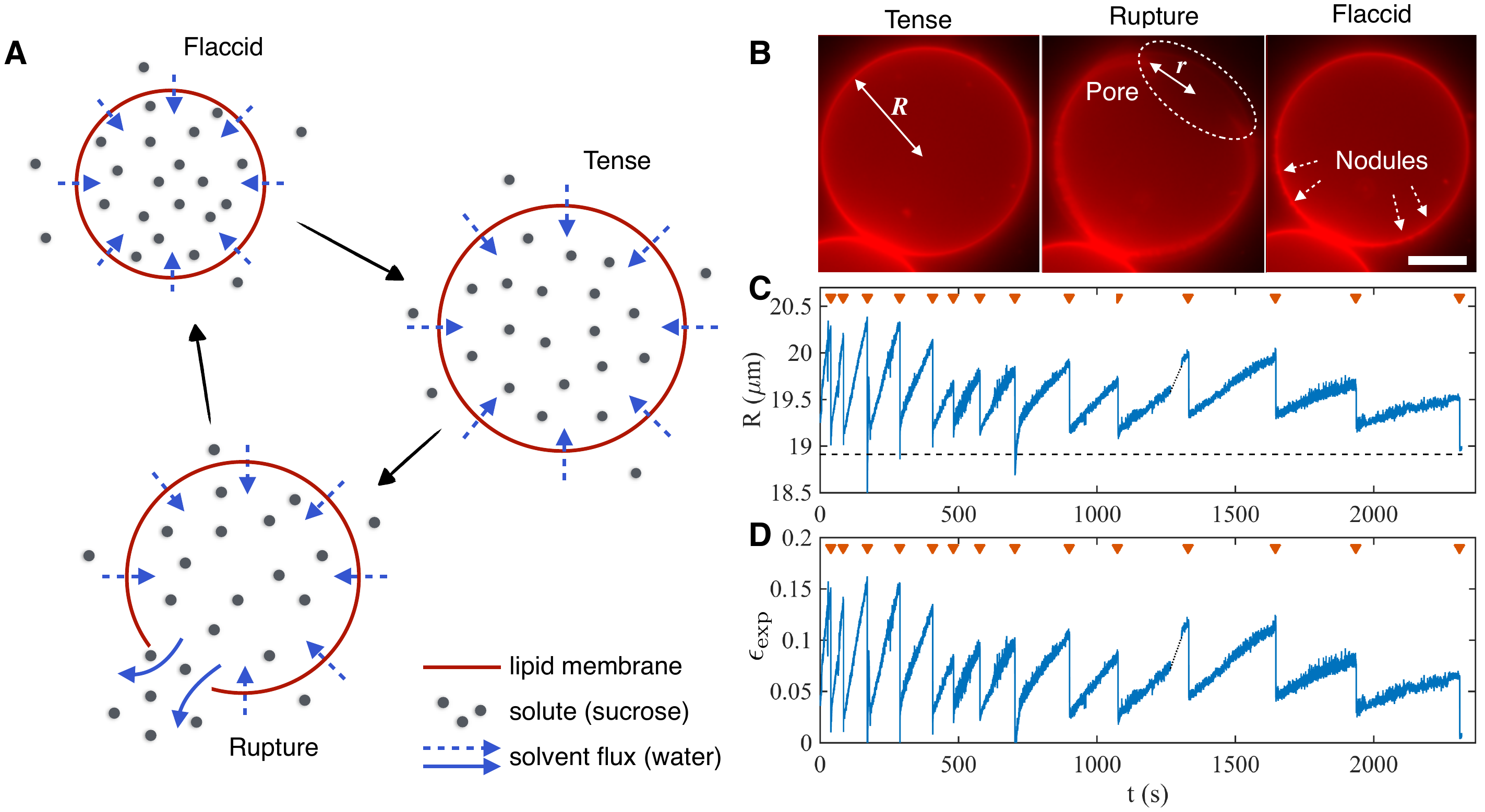}
\caption{Homogeneous giant unilamellar vesicles (GUVs) made of POPC with 1 mol $\%$ Rho-DPPE exhibit swell-burst cycles when subject to hypotonic conditions.  (A) Schematic of a swell-burst cycle of a homogeneous GUV under hypotonic conditions. Blue arrows represent the leak-out of the inner solution through the transient pore. (B) Micrographs of a swelling (left), ruptured (middle) and resealed (right) GUVs. Scale bar represents 10 $\mu$m. Pictures extracted from Movie~S1 in the Supporting Material. (C) Typical evolution of a GUV radius with time during swell-burst cycles in 200~mM sucrose hypotonic conditions. The GUV radius increases continuously during swelling phases, and drops abruptly when bursting events occur. Pore opening events are indicated by red triangles. Dashed line represents the estimated initial radius $R_0$. See also Movie~S2 in the Supporting Material. More GUV radius measurements are shown in Fig.~S\ref{figS_exp} in the Supporting Material. (D) Experimental area strain \revise{$\epsilon_\text{exp}$ as a function of time}.  }
\label{fig1}
\end{figure*}

A selection of snapshots, revealing different morphological states, and a detailed trace showing \revise{the time-dependence of the vesicle radius $R$ and corresponding area strain ($\epsilon_\text{exp}=(R^2-R_0^2)/R_0^2$, where and $R_0$ is the resting initial vesicle radius) are shown} in Fig.~\ref{fig1}(B, C, and D), for a representative GUV. Swelling phases are characterized by a quasi-linear increase of the GUV radius, while pore openings cause a sudden decrease of the vesicle radius. 

We outline here three key observations about the dynamics of swell-burst cycles from these experiments. 
\begin{enumerate}
\item The period between two consecutive bursting events increases with each cycle, starting from a few tenths of a second for the early cycles, to several hundreds of seconds after the tenth cycle.
\item The maximum radius and therefore the maximum strain at which a pore opens decreases with cycle number, suggesting that lytic tension is a dynamic property of the membrane.
\item The observed transient pores are short lived, stay open for about a hundred milliseconds, and reach a maximum radius of up to 60~$\%$ of the GUV radius.
\end{enumerate}

We seek to explain these observations through a quantitative understanding of the pulsatile GUVs in hypotonic conditions. To do so, we first investigate the mechanics of pore nucleation and its relationship to the GUV swell-burst dynamics.

\subsection*{\hf Thermal fluctuations drive the dynamics of pore nucleation}

\begin{figure*}[tb] 
\centering
\includegraphics[scale=0.5]{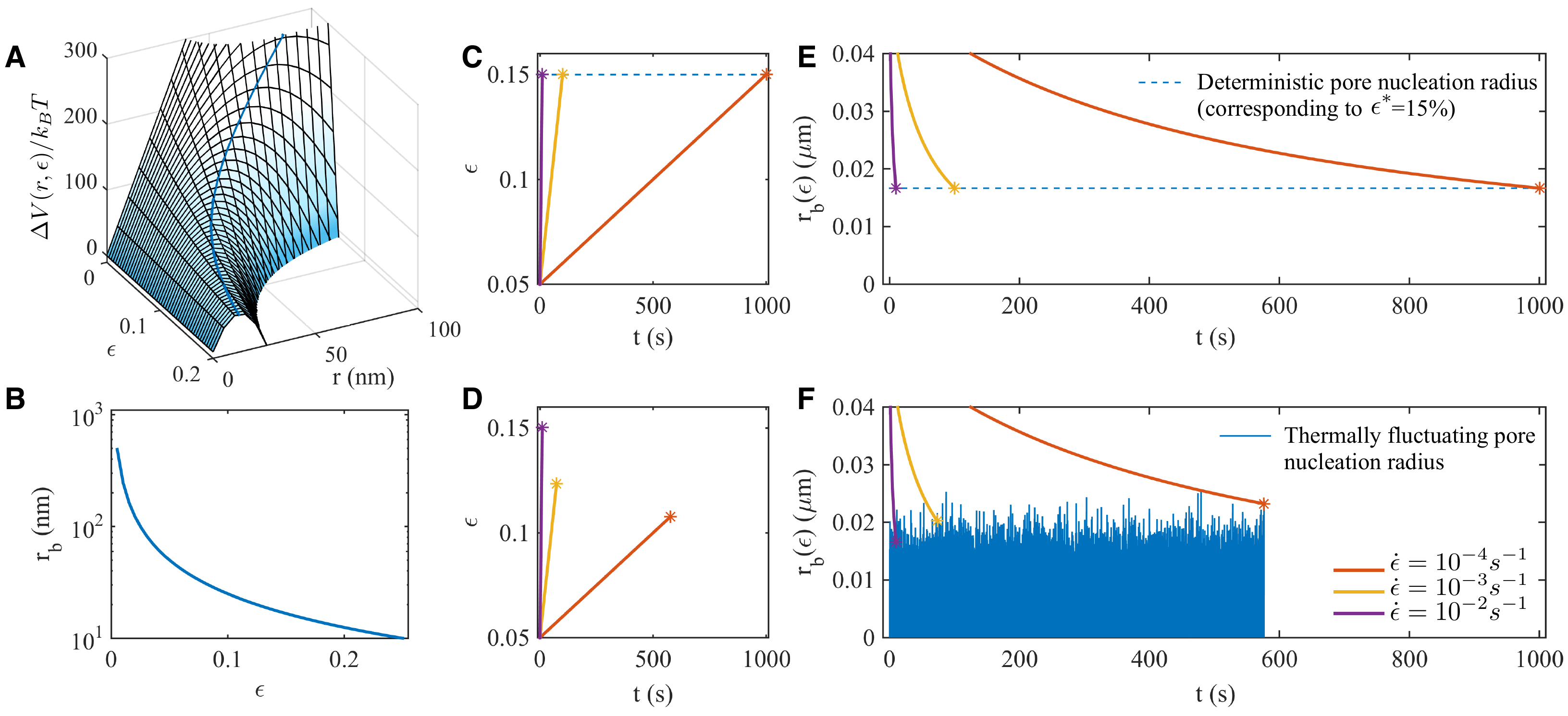}
\caption{Lytic tension is a dynamic quantity governed by thermodynamic fluctuations. (A) The energy required to open a pore of radius $r$ in a GUV without fluctuations, for various membrane tensions. The energetic cost to open a pore in a tense GUV shows a local maximum, which has to be overcome in order for a pore to open. (B) For a given strain $\epsilon$, the energy barrier is located at a pore radius \revise{$r_b=\gamma/\sigma(\epsilon)=\gamma/(\kappa_\text{eff}\epsilon)$ (where $\kappa_\text{eff}=2\times 10^{-3}$ N/m as discussed in the text)}. (C and E) If the critical strain is fixed at a constant value, $\epsilon^*$, as in the deterministic approach, then a pore is nucleated whenever the strain reaches $\epsilon^*$, regardless of the the strain rate. Prescribing various linear strain rate ($\epsilon = \dot{\epsilon}t + \epsilon_0$, with $\dot{\epsilon}=10^{-2}$, $10^{-3}$, and $10^{-4}$ s$^{-1}$, $\epsilon_0=0.05$) does not alter the strain at which a pore will form (C) and the pore nucleation radius $r_b$ will be constant (E). (D and F) In the stochastic approach however, the nucleation threshold is replaced by a fluctuating pore \revise{(blue line in (F) as computed by Eq.~\eqref{eq:sde_r})}, inducing a dependence of the lytic strain on the strain rate (D). This is due to the fact that, for lower strain rates, the probability of a large pore fluctuation to reach $r_b$ is higher (F), producing a lower lytic tension on average.}
\label{fig2}
\end{figure*}

\revise{
In the framework of classical nucleation theory \cite{litster1975}, the energy potential $V(r,\epsilon)$ of a pore of radius $r$ in a lipid membrane under surface tension $\sigma$, is the balance of two competitive terms: $V_s(\epsilon)$, the strain energy, and $V_p(r)$, the pore energy. The strain energy tends to favor the opening and enlargement of the pore while the pore closure is driven by the pore line tension $\gamma$. Accordingly, the energy potential reads}

\begin{align}
 V(r,\epsilon)=&V_s(\epsilon) + V_p(r) \nonumber\\
 =&\dfrac{1}{2}\kappa_\text{eff} A_0 \epsilon^2 + 2\pi r \gamma \;.
 \end{align}
  
\revise{The area strain is defined as $\epsilon=(A-A_0)/A_0$, where $A=4\pi R^2-\pi r^2$ is the surface of the membrane, and $A_0=4\pi R_0^2$ is the resting vesicle area.
Here $V_s(\epsilon)$ is assumed to have a Hookean form, where $\kappa_\text{eff}$ is the effective stretching modulus, which relates the surface tension to the strain as $\sigma=\kappa_\text{eff}\epsilon$ (see next section for a discussion on $\kappa_\text{eff}$).
These two energetic terms oppose each other, resulting in an energy barrier that the system has to overcome in order for a pore to nucleate. The competition between the strain and pore energy is expressed by the ratio $r_b=\gamma/\sigma$, which is the critical radius associated with the crossing of the energy barrier. That is, if a pore in a tensed membrane has a radius $r<r_b$, the pore energy $V_p(r)$ dominates and the pore closes. On the contrary, for $r>r_b$, the strain energy $V_s(\epsilon)$ prevails and the pore grows.
The energy required to open a pore of radius $r$ in a tensed GUV is given by $\Delta V(r,\epsilon) =  V(r,\epsilon)-V(0,\epsilon)$ and is represented in Fig.~\ref{fig2}(A). The corresponding critical radius of the the energy barrier $r_b$ is shown as a function of the strain $\epsilon$ in Fig.~\ref{fig2}(B). The height of the energy barrier and its critical radius are dependent on the membrane strain; the more the membrane is stretched, the lower the energy barrier is, and the smaller the amount of energy required to nucleate a pore. }

The amplitude of this energy barrier is strictly positive for finite strain values, making pore nucleation impossible without the addition of external energy. This issue has been often resolved by assuming a predetermined and \textit{constant} lytic \revise{strain ($\epsilon^*$)} corresponding to a critical energy barrier under which the pore opens (Fig.~\ref{fig2}(C and E)). However, this approach is in contradiction with our experimental observations that the lytic strain in the membrane varies with each swell-burst cycle (Fig.~\ref{fig1}(D)), due to a dependence on the strain rate \cite{evans2003}. In order to account for this variation, we included thermal fluctuations associated with the pore nucleation barrier in our analysis \cite{ting2011, bicout2012}. In this scenario, increasing the membrane tension of the vesicle reduces the minimum pore radius $r_b$ at which a pore opens (Fig.~\ref{fig2}(A and B)), lowering the energy barrier down to the range of thermal fluctuations, eventually letting the free energy of the system to overcome the nucleation barrier (Fig.~\ref{fig2}(D and F)). The stochastic nature of the fluctuations can then explain a distribution of pore opening tensions, eliminating the need to assume constant lytic tension. 

A direct consequence of the fluctuation-mediated pore nucleation is that the membrane rupture properties become dynamic. Indeed, fluctuations naturally cause the strain at which the membrane ruptures to be dependent on the \textit{strain rate}, as illustrated in Fig.~\ref{fig2}(D). In order to understand this dynamic nucleation process, consider stretching the membrane at different strain rates $\dot{\epsilon}$. Doing so decreases the radius of the nucleation barrier at corresponding speeds, as shown in Fig.~\ref{fig2}(F). For slow strain rates, as $r_b$ tends to zero, it spends more time in the accessible range of the thermal pore fluctuations, increasing the probability that a fluctuation will overcome the energy barrier. On the other hand, at faster strain rates, $r_b$ decreases quickly, reaching small values in less time, lowering the probability for above average fluctuations to occur during this shorter time.
 
We use a Langevin equation to capture the stochastic nature of pore nucleation and the subsequent pore dynamics. This equation includes membrane viscous dissipation, a conservative force arising from the membrane potential, friction with water, and thermal fluctuations for pore nucleation (see Model Development and Simulations in the Supporting Material for detailed derivation). This yields the stochastic differential equation for the pore radius $r$
\begin{equation}\label{eq:sde_r}
\overbrace{\left(h \eta_m + C \eta_s r\right)}^{\mathclap{\text{viscous drag}}} 
\underbrace{\frac{d}{dt}(2\pi r)}_{\mathclap{\text{change of pore radius}}}  = 
\overbrace{2\pi \left(\sigma r-\gamma\right)}^{\mathclap{\substack{\text{surface and}\\\text{line tension}}}} + 
\underbrace{\xi(t)}_{\mathclap{\substack{\text{thermal}\\\text{pore fluctuations}}}}  \;,
\end{equation}
where the noise source $\xi(t)$ has zero mean and satisfies, $\langle \xi(t)\xi(t^\prime) \rangle = 2\left(h \eta_m + C \eta_s r\right) k_B T \delta(t-t^\prime)$ according to the fluctuation dissipation theorem \cite{kubo1966}. Here, $\eta_m$ and $\eta_s$ are the membrane and solute viscosities respectively, $h$ is the membrane thickness, $C$ is a geometric coefficient \cite{ryham2011, aubin2016}, $k_B$ is the Boltzmann constant and $T$ is the temperature. \revise{We assume here that the pore nucleation probability is independent on the total membrane surface area.} The values of the different parameters used in the model are given in Table~S\ref{tab_parameters} in the Supporting Material.


\subsection*{\hf Model captures experimentally observed pulsatile GUV behavior}

\begin{figure}[tb] 
\centering
\includegraphics[scale=0.55]{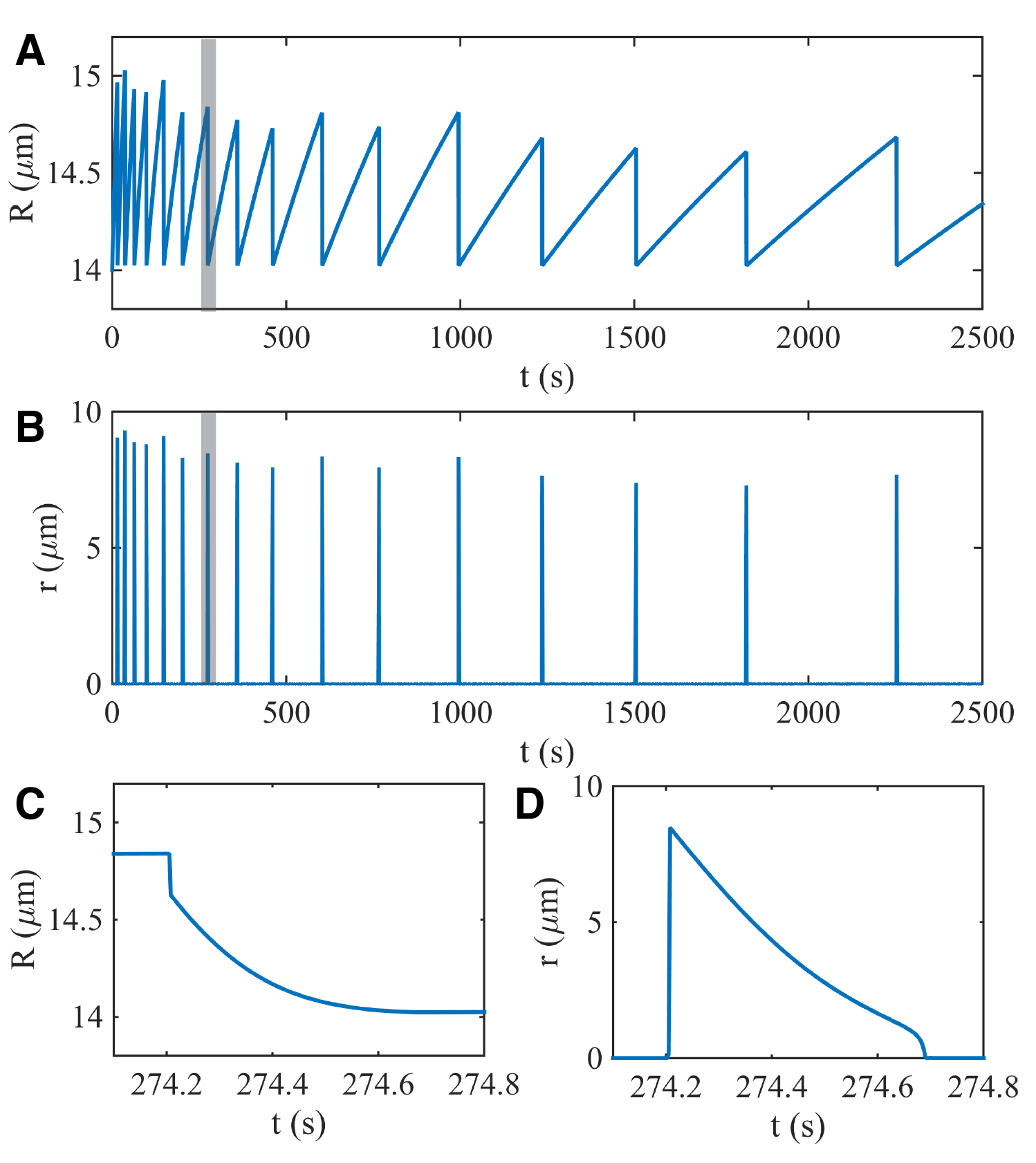}
\caption{Dynamics of swell-burst cycles from the model for a GUV of radius 14~$\mu$m in 200 mM hypotonic stress. (A and C) GUV radius and (B and D) pore radius as a function of time. The model captures the dynamics of multiple swell-burst cycles, in particular the decrease of maximum GUV radius and increase of cycle period with cycle number (A). Looking closely at a single pore opening event corresponding to the grey region, the model predicts three stage pore dynamics (C and D), namely opening, closing, and resealing, with a characteristic time of a few hundred milliseconds. Numerical reconstruction of the GUV is shown in Movies~S3 and~S4 in the Supporting Material. Results for $R_0=8$ and 20~$\mu$m are shown in Fig.~S\ref{figS_R} in the Supporting Material.}
\label{fig3}
\end{figure}

In addition to pore dynamics (Eq.~\ref{eq:sde_r}), we need to consider mass conservation of the solute and the solvent. We assume that the GUV remains spherical at all times and neglect spatial effects. The GUV volume changes because of osmotic influx through the semi-permeable membrane and the leak-out of the solvent through the pore. The osmotic influx is the result of two competitive pressures, the osmotic pressure driven by the solute differential ($\Delta p_{osm} = k_B T N_A \Delta c$), and the Laplace pressure, arising from the membrane tension ($\Delta p_L = 2\sigma / R$), resulting in the following equation for the GUV radius $R$:
\begin{equation} \label{eq:ode_R}
\underbrace{\frac{d}{dt}\left( \frac{4}{3}\pi R^3 \right)}_{\mathclap{\text{change of GUV volume}}}
 = \overbrace{\frac{P \nu_s }{k_B T N_A} \left( \Delta p_{osm} - \Delta p_L \right) A}^{\substack{\text{influx of solvent}\\\text{ through the membrane} }}
 - \underbrace{ v_L \pi r^2}_{\mathclap{\substack{\text{leak-out}\\\text{through the pore}}}} \;.
\end{equation}
Here $A=4\pi R^2$ is the membrane area, $P$ is the membrane permeability to the solvent, $\nu_s$ is the solvent molar volume, and $N_A$ is the Avogadro number.  Assuming low Reynolds number regime, the leak-out velocity is given by $ v_L = \Delta p_L r / (3\pi\eta_s)$ \cite{happel1983, aubin2016}.

Mass conservation of solute in the GUV is governed by the diffusion of sucrose and convection of the solution through the pore, which gives the governing equation for the solute concentration differential $\Delta c$:
\begin{equation} \label{eq:ode_c}
\underbrace{\frac{d}{dt}\left(\frac{4}{3}\pi R^3 \Delta c \right)}_{\mathclap{\substack{\text{molar differential}\\\text{of solute}}}}
 = -\pi r^2 \bigg( \overbrace{D \frac{\Delta c}{R}}^{\mathclap{\substack{\text{diffusion through}\\\text{the pore}}}}
  + \underbrace{v_L\Delta c}_{\mathclap{\substack{\text{convection}\\\text{through the pore}}}} \bigg)  \;,
\end{equation}
where $D$ is the solute diffusion coefficient. These three coupled equations (Eqs.~\ref{eq:sde_r} to \ref{eq:ode_c}) constitute the mathematical model. 

In order to completely define the system, we need to specify the relationship between the membrane surface tension $\sigma$ and the area strain of the GUV. We note that the GUV has irregular contours during the pore opening event and for a short time afterwards, when ``nodules"  are observed at the opposite end from the pore, indicating accumulation of excess membrane generated by pore formation (Fig.~\ref{fig1}(B) middle and right panels). In the low tension regime, GUVs swell by unfolding these membrane nodules, and the stretching is controlled by the membrane bending modulus $\kappa_b$ and thermal energy, yielding an effective ``unfolding modulus" $\kappa_u = 48\pi\kappa_b^2 / (R_0^2k_BT)$ of the order of 10$^{-5}$ N/m \cite{brochard1976}.  In contrast, in the high tension regime, elastic stretching is dominant, and the elastic area expansion modulus $\kappa_e$ is roughly equal to 0.2 N/m \cite{evans1990}. Since the maximum area strain plotted in Fig.~\ref{fig1}(D), is about 15~$\%$, significantly larger than the expected 4~$\%$ for a purely elastic membrane deformation, the experimental data suggests the occurrence of two stretching regimes: an unfolding driven stretching, and an elasticity driven stretching \cite{ertel1993, hallett1993, karatekin2003a}. \revise{Therefore, for simplicity, we assume an effective stretching modulus $\kappa_\text{eff}$, which takes into account both unfoldoing and elastic regimes \cite{evans1990, bloom1991} through a linear dependence between the membrane tension and the strain ($\sigma = \kappa_\text{eff} \epsilon$). Note that $\kappa_\text{eff}$ is the only adjustable parameter of the model.}

We solved the three coupled equations (Eqs.~\ref{eq:sde_r} to \ref{eq:ode_c}) for an initial inner solute concentration of $c_0=200$~mM, and different GUV radii of $R_0=8$, 14 and 20~$\mu$m. All the results presented here are obtained  for $\kappa_\text{eff}=2 \times10^{-3}$N/m, the value that best fits the experimental observations (see Supplemental Fig.~\ref{figS_kappa} for the effect of this parameter on the GUV dynamics). Dynamics of the GUV radius and the pore radius are shown in Fig.~\ref{fig3} for a typical simulation with $R_0=14$~$\mu$m (see Supplemental Fig.~\ref{figS_R} for simulations with different values of $R_0$). Our model qualitatively reproduces the dynamics of the GUV radius during the swell burst cycle (compare Figs.~\ref{fig1}(C) and \ref{fig3}(A)). Importantly, we recover the key features of the swell-burst cycle -- namely an increase of the cycle period with each bursting event (point 1), and a decrease of the maximum radius with time (point 2). The stochastic nature of the thermodynamic fluctuations leads to variations and irregularities in the pore opening events, and therefore, the cycle period and maximum strain. The dynamics of a single cycle is shown in Fig.~\ref{fig3}(C and D). Our numerical results show an abrupt drop in the GUV radius, followed by a slower decrease, suggesting a sequence of two leak-out regimes: a fast-burst releasing most of the membrane tension, and a low tension leak-out. This two-step tension release is confirmed by the pore radius dynamics, which after suddenly opening (release of membrane tension), reseals quasi-linearly due to dominance of line tension compared to membrane tension in Eq.~\ref{eq:sde_r}. Furthermore, the computed pore amplitude and lifetime are in agreement with experimental observations (point 3). Overall, our model is able to reproduce the quantitative features of GUV response to hypotonic stress over multiple time scales.

If thermal fluctuations are ignored, the strain to rupture needs to be adjusted to roughly 15$\%$ in order to match the range of maximum GUV radius observed experimentally (Fig.~S\ref{figS_det} in the Supporting Material). However such a deterministic model does not capture the pulsatile dynamics as well as the stochastic model in terms of cycle period and strain rate (Fig.~S\ref{figS_comparison} in the Supporting Material), and fails to reproduce a strain rate dependent maximum stress (Fig.~S\ref{figS_det}).

\subsection*{\hf Solute diffusion is dominant during the low tension regime of pore resealing}

\begin{figure}[tbp] 
\centering
\includegraphics[scale=0.55]{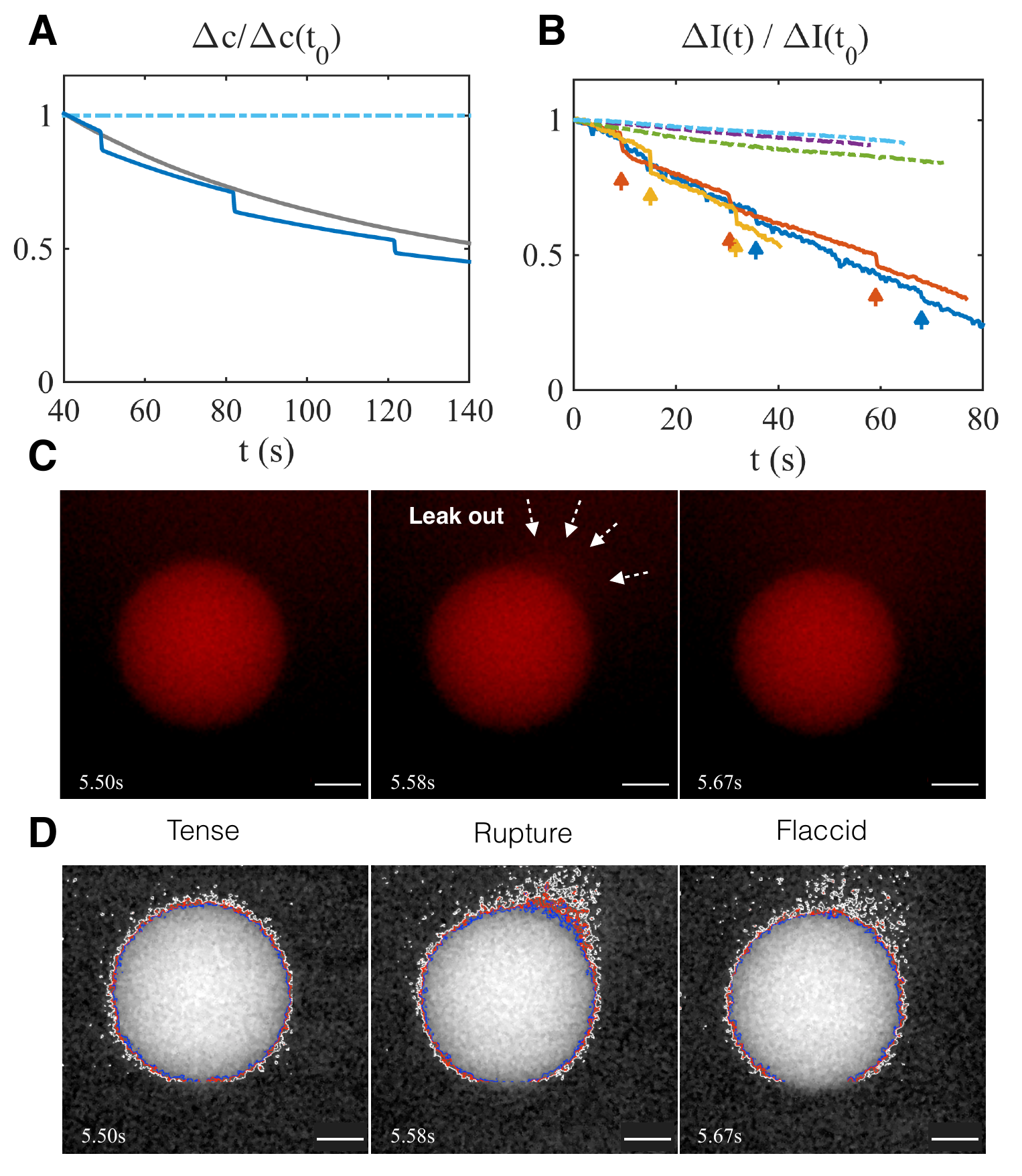}
\caption{\revise{Diffusion of sucrose through the transient pore produces a step-wise decrease of the inner solute concentration.
(A) In hypotonic conditions, the model predicts a step-wise decrease of solute concentration differential with time (blue line), which is solely due to diffusion of solute through the transient pore. In comparison, when diffusion is neglected in the model (grey line), the solute concentration differential decreases smoothly  (also see Fig.~S\ref{figS_D} in the Supporting Material for further analysis on the  effect of diffusion). In isotonic conditions (dashed line), the solute concentration differential is constant with time. (Here $t_0=$ 40 s).}
(B) Time evolution of the normalized fluorescence intensity of a GUV in hypotonic condition, encapsulating fluorescent glucose analog. $\Delta I$ is the difference in mean intensity between the inside of the GUV and the background. In hypotonic conditions (solid lines) the normalized intensity decreases with time due to the constant influx of water through the membrane, and shows sudden drops in intensity at each pore opening (indicated by arrows), due to diffusion of sucrose through the pore (see Movie~S5 in the Supporting Material). In comparison, GUVs in an isotonic environment (dashed lines) exhibit a rather constant fluorescence intensity (see Movie~S6 in the Supporting Material). (C) Micrographs of a GUV in hypotonic condition, encapsulating fluorescent glucose analog, just prior to bursting (left panel), with an open pore (middle panel), and just after pore resealing (right panel). The leak-out of fluorescent dye is observed in the middle frame, coinciding with a drop of the GUV radius. Frames extracted from Movie~S7 in the Supporting Material. (D) Same as panel (C), with the images processed to increase contrast and attenuate noise. The blue, red, and white lines are the isocontours of the 90, 75, and 60 grey scale values respectively, highlighting the leak-out of fluorescent dye.}
\label{fig4}
\end{figure}

The concentration differential of sucrose decreases exponentially and drops from 200 mM to about 10 mM in about 1000 seconds (Fig.~S\ref{figS_R} in the Supporting Material). Even after 2000 s when the concentration differential is as low as 10~mM, the osmotic influx is still large enough to maintain the dynamics of swell-burst cycles (Fig.~\ref{fig1}(C), Fig.~S\ref{figS_R}). We further observe that every pore opening event produces a sudden drop in inner solute concentration (Fig.~\ref{fig4}(A), blue line). This suggests that diffusion of sucrose plays an important role in governing the dynamics of solute. In the absence of diffusive effects, the model does not show the abrupt drops in concentration but a rather smooth exponential decay (Fig.~\ref{fig4}(A), grey line). 

To experimentally verify the model predictions of sucrose dynamics, we quantified the evolution of fluorescence intensity in GUVs encapsulating 200 mM sucrose plus 58.4 $\mu$M 2-NBDG, a fluorescent glucose analog (see Supporting Material and Methods). Fig.~\ref{fig4}(B) presents the evolution of fluorescent intensity of sucrose in time. GUVs in isotonic conditions (dashed lines) do not show a significant change in fluorescence intensity. GUVs in hypotonic conditions (solid lines) exhibit an overall decrease of intensity due to permeation of water through the membrane. Strikingly, consecutive drops of fluorescence intensity are observed coinciding with the pore opening events (Fig.~\ref{fig4}(C and D) middle panels), and point out the importance of sucrose diffusion through the pore. While the quantitative dynamics of sucrose depends on the value of the diffusion constant (Fig.~S\ref{figS_D}), the qualitative effect of diffusion on the dynamics remains unchanged. On the other hand, leak-out induced convection does not influence the inner concentration of sucrose, as both solvent and solute are convected, conserving their relative amounts. These observations are in agreement with the existence of the low tension pore closure regime discussed above, where Laplace pressure produces negligible convective transport compared to solute diffusion though the pore.

\subsection*{\hf Cycle period and strain rate are explicit functions of the cycle number and GUV properties}

%

\begin{figure*}[tb] 
\centering
\includegraphics[scale=0.55]{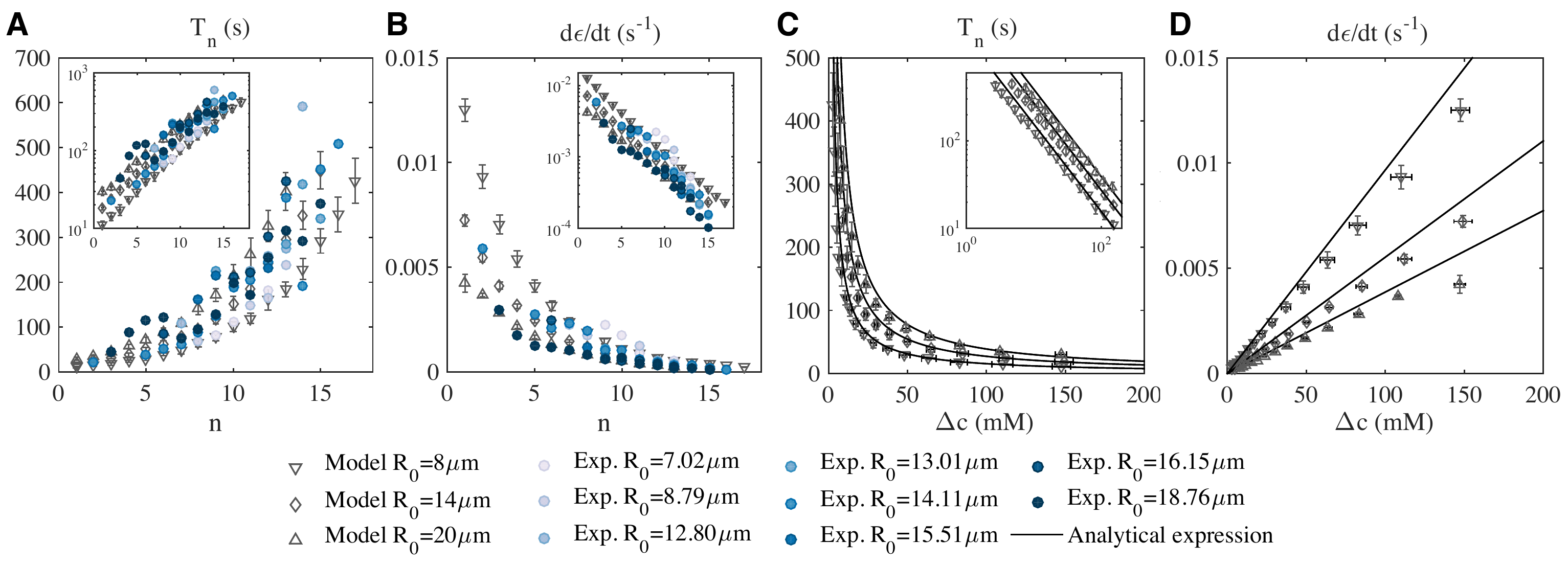}
\caption{Cycle period and strain rate are exponential functions of cycle number, and power-law functions of solute concentration.
 (A, C) Cycle period and (B, D) strain rates as functions of cycle number ($n$) (A, B) and solute concentration (C, D).  Insets show the same data in log scale. Each model point is the mean of 10 numerical experiments, error bars represent $\pm$ standard deviations. The analytical expressions for the \revise{cycle period and strain rate (Eq.~\eqref{eq:cycle_period})} with $\epsilon^*=0.15$, are plotted in (C, D) for comparison.
}
\label{fig5_1}
\end{figure*}

\begin{figure*}[tb] 
\centering
\includegraphics[scale=0.55]{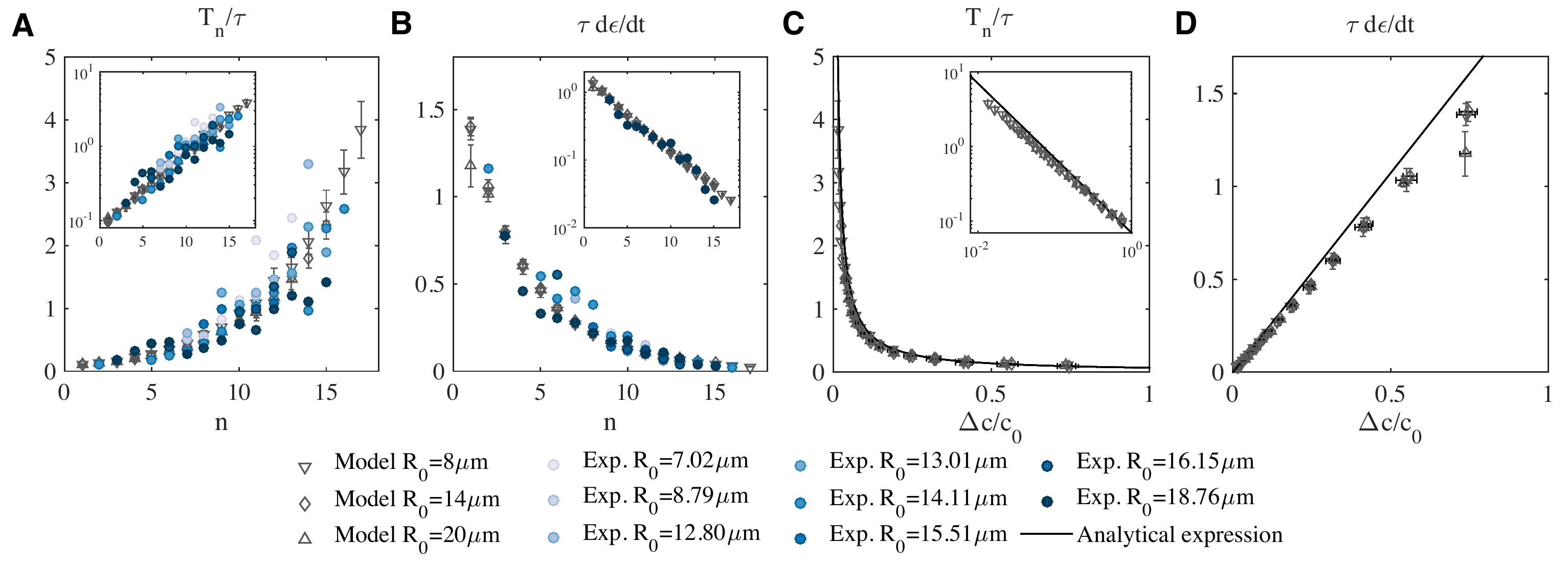}
\caption{The pulsatile dynamics is characterized by the characteristic time-scale $\tau$. The data from Fig.~\ref{fig5_1} are scaled by the characteristic time associated with swelling defined as \revise{$\tau$}. Insets show the same data in log scale.  The non-dimensionalization by $\tau$ allows cycle periods and the strain rates to collapse onto a single curve. (C, D) The analytical expressions for the \revise{cycle period and strain rate (Eq.~\eqref{eq:cycle_period})} with $\epsilon^*=0.15$, are plotted for comparison.
}
\label{fig5_2}
\end{figure*}

Given that lytic tension is a dynamic quantity, we asked how cycle period and strain rate evolve along with the cycles. We analyzed the simulated dynamics of GUVs with resting radii of 8, 14 and 20$\mu m$, each data point representing the mean and the standard deviation of 10 simulations with identical parameters (the variations being due to the stochastic nature of the model). \revise{The details of this burst cycle analysis is reported in the Supporting Material}. Cycle periods and strain rates show a dependence on the GUV radius, as depicted in Fig.~\ref{fig5_1} where larger GUVs have slower dynamics, resulting in smaller strain rates and longer cycle periods (Fig.~S\ref{figS_R}). To verify this experimentally, a total of eight GUVs were similarly analyzed with resting radii ranging from 7.02 to 18.76 $\mu$m (Fig.~S\ref{figS_exp} in the Supporting Material). The measured cycle period and strain rate as a function of the cycle number (corrected for the lag between the application of the hypotonic stress and the beginning of the observations) are shown in Fig.~\ref{fig5_1}(A) and (B), respectively. Experimental and model results quantitatively agree, and show a exponential dependence of the cycle period and strain rate on cycle number (Insets Fig.~\ref{fig5_1}(A and B)). 

Two further questions arise: How can we relate the cycle number to the driving force of the process, namely the osmotic differential? And, is there a scaling law that governs the GUV swell-burst dynamics? To answer these questions we computed the cycle solute concentration (defined as the solute concentration at the beginning of each cycle) as a function of the cycle number (Fig.~S\ref{figS_cn} in the Supporting Material). We found that the solute concentration follows an exponential decay function of the cycle number, and is independent of the GUV radius. Additionally, plotting the cycle period and strain rate against the cycle solute concentration (Fig.~\ref{fig5_1}(C and D)), we observe that the cycle period increases as $\Delta c$ decreases, while the strain rate is a linear function of $\Delta c$. The data presented in Fig.~\ref{fig5_1} suggest that the dynamics of GUVs swell-burst cycle can be scaled to their size. From the non-dimensional form of Eq.~\ref{eq:ode_R}, we extracted a characteristic time associated with swelling, defined by $\tau = R_0 / (P \nu_s c_0)$, and scaled the cycle period and strain rates with this quantity. As shown in Fig.~\ref{fig5_2}, all the scaled experimental and model data collapse onto the same curve, within the range of the standard deviations. The scaled relationships can be justified analytically, by estimating the cycle period and strain rates as \revise{
\begin{equation} \label{eq:cycle_period}
\frac{ T_n}{\tau} \simeq \frac{\epsilon^*}{\left( 2\sqrt{\epsilon^*+1} \Delta c/c_0 \right)}
 \qquad  \text{and} \qquad
 \tau \dot{\epsilon} \simeq \frac{2\sqrt{\epsilon^*+1} \Delta c}{c_0} 
\end{equation} }
respectively (see Supporting Material for full derivation). These analytical expressions are plotted in Figs.~\ref{fig5_1}(C, D) and \ref{fig5_2}(C, D) for a characteristic lytic strain of $\varepsilon^*=0.15$, showing good agreement with the numerical data. Taken together, these results suggest that the GUV pulsatile dynamics is governed by the radius, the membrane permeability, the solute concentration, and importantly the stochastic pore nucleation mechanism which determines the strain to rupture.

\section*{\hf DISCUSSION}

Explaining how membrane-enclosed compartments regulate osmotic stress is a first step towards understanding how cells control volume homeostasis in response to environmental stressors. In this work, we have used a combination of theory, computation, and experiments in a simple model system to study how swell-burst cycles control the dynamics of GUV response to osmotic stress. Using this system, we show that the pulsatile dynamics of GUVs under osmotic stress is controlled through thermal fluctuations that govern pore nucleation and lytic tension.

The central feature of a GUV's osmotic response is the nucleation of a pore.  Even though Evans and coworkers \cite{evans2003, evans2011} identified that rupture tension was not governed by an intrinsic critical stress, but rather by the load rate, the idea of a constant lytic tension has persisted in the literature \cite{idiart2004, popescu2008, peterlin2008}. By coupling fluctuations to pore energy, we have now reconciled the dynamics of the GUV over several swell-burst cycles with pore nucleation and dependence on strain rate. Our model is not only able to capture the experimentally observed pulsatile dynamics of GUV radius and solute concentration (Figs.~\ref{fig3} and \ref{fig4}), but also predicts pore formation events and pore dynamics (Fig.~\ref{fig3}(B and D)). We also found that during the pore opening event, a low-tension regime enables a diffusion dominated transport of solute through the pore (Fig.~\ref{fig4}), a feature that has been until now neglected in the literature.

Specifically, we have identified a scaling relationship between (a) the cycle period and cycle number and (b) the strain rate and the cycle number, highlighting that swell-burst cycles of the GUVs in response to hypotonic stress is a dynamic response (Fig.~\ref{fig5_2}). One of the key features of the model is that we relate the cycle number, an experimentally observable quantity, to the concentration difference of the solute, a quantity that is hard to measure in experiments (Fig.~S\ref{figS_cn}). This allows to interpret the scaling relationships described above in terms of solute concentration differential. The cycle period increases as the solute concentration difference decreases, while the strain rate is a linear function of the concentration difference. Both relationships are derived theoretically in the Supplemental material. These features indicate long time scale relationships of pulsatile vesicles in osmotic stress.

Thermal fluctuations and stochasticity are known to play diverse roles in cell biology. Well-recognized examples include Brownian motors and pumps \cite{julicher1997, oster2002}, noisy gene expression \cite{elowitz2002}, and red blood cell flickering \cite{turlier2016}. The pulsatile vesicles presented here provide yet another example of how fluctuations can be utilized by simple systems to produce dynamical adaptive behavior. Given the universality of fluctuations in biological processes, it appears entirely reasonable that simple mechanisms similar to these pulsatile vesicles may have been exploited by early cells, conferring them with a thermodynamic advantage against environmental osmotic assaults. On the other hand, if such swell-burst mechanisms were at play, the chronic leak-out of inner content could have led protocells to evolve active transport mechanisms to compensate for volume loss, and endure osmotic stress without a high energetic cost.

In this study, we experimentally measure the dynamics of swell-burst cycles in GUVs, and provide for the first time a model that captures quantitatively the pulsatile behavior of GUVs under hypotonic conditions for long time scales. In order to do so, we developed a general framework which integrated parts of existing models \cite{koslov1984, brochard-wyart2000, ryham2011}, with novel key elements: (a) the explicit inclusion of thermal pore fluctuations, which enables dynamic pore nucleation; (b) the definition of an effective stretching modulus, which combines membrane unfolding and elastic stretching; (c) the incorporation of solute diffusion through the pore, which results in a non-trivial contribution to the evolution of the osmotic differential. The coupling of these key features results in a unified model that is valid in all regimes of the vesicle, pore, and solute dynamics.

While we have been able to explain many fundamental features of the pulsatile GUVs in response to osmotic stress, our approach has some  limitations and there is a need for further experiments. We have assumed a linear relationship between stress and strain. Although this assumption is reasonable and appears to work well for the present experimental conditions, a more general expression should be considered to include both membrane (un)folding and elastic deformation \cite{helfrich1984}. 
Another important aspect of biological relevance is membrane composition, where the abundance of proteins and heterogeneous composition leading to in-plane order and asymmetry across leaflets influence the membrane mechanics \cite{alberts2014, rangamani2014}. We have previously found experimentally that the dynamics of swell-burst cycles is related to the compositional degrees of freedom of the membrane \cite{oglecka2014}. Future efforts will be oriented toward the development of theoretical framework and quantitative experimental measures that provide insight into how the membrane's compositional degrees of freedom influence the pulsatile dynamics of cell-size vesicles. \revise{In addition to osmotic response and membrane composition, we will focus on how membrane components such as aquaporins and ion channels may couple thermal fluctuations with membrane tension to regulate their functions. Additionally, we are also investigating how the properties of the encapsulated bulk fluid phase may affect the response of the GUV in response to osmotic shock. The current work is a first and critical step in these directions. }


\section*{\hf AUTHOR CONTRIBUTIONS}
J.C.S.H. and A.N.P. designed the experiments;
J.C.S.H. performed the experiments;
M.C. and P.R. derived the model;
M.C. performed the simulations;
M.C. and J.C.S.H. analyzed the data;
all authors discussed and interpreted results; all authors wrote and agreed on the manuscript.

\section*{\hf ACKNOWLEDGMENTS}
We are grateful to Prof. Wouter-Jan Rappel and Prof. Alex Mogilner for insightful comments on the manuscript. We also thank Prof. Daniel Tartakovsky for enriching discussions. This work was supported in part by the FISP 3030 for the year 2015-2016 to M.C., NTU provost office to J.C.S.H., AFOSR FA9550-15-1-0124 award to P.R., and NSF PHY-1505017 award to P.R. and A.N.P.

\bibliographystyle{biophysj}
\bibliography{biblio}

\begin{thebibliography}{41}
\providecommand{\url}[1]{\texttt{#1}}
\providecommand{\urlprefix}{ }

\bibitem[Christensen(1987)]{christensen1987}
Christensen, O., 1987.
\newblock Mediation of cell volume regulation by Ca2+ influx through
  stretch-activated channels.
\newblock \emph{Nature} 330:66--68.

\bibitem[Hoffmann et~al.(2009)Hoffmann, Lambert, and Pedersen]{hoffmann2009}
Hoffmann, E.~K., I.~H. Lambert, and S.~F. Pedersen, 2009.
\newblock Physiology of {{Cell Volume Regulation}} in {{Vertebrates}}.
\newblock \emph{Physiological Reviews} 89:193--277.

\bibitem[Berrier et~al.(1996)Berrier, Besnard, Ajouz, Coulombe, and
  Ghazi]{berrier1996}
Berrier, C., M.~Besnard, B.~Ajouz, A.~Coulombe, and A.~Ghazi, 1996.
\newblock Multiple mechanosensitive ion channels from Escherichia coli,
  activated at different thresholds of applied pressure.
\newblock \emph{The Journal of membrane biology} 151:175--187.

\bibitem[Blount et~al.(1997)Blount, Schroeder, and Kung]{blount1997}
Blount, P., M.~J. Schroeder, and C.~Kung, 1997.
\newblock Mutations in a bacterial mechanosensitive channel change the cellular
  response to osmotic stress.
\newblock \emph{Journal of Biological Chemistry} 272:32150--32157.

\bibitem[Levina et~al.(1999)Levina, T{\"o}temeyer, Stokes, Louis, Jones, and
  Booth]{levina1999}
Levina, N., S.~T{\"o}temeyer, N.~R. Stokes, P.~Louis, M.~A. Jones, and I.~R.
  Booth, 1999.
\newblock Protection of Escherichia coli cells against extreme turgor by
  activation of MscS and MscL mechanosensitive channels: identification of
  genes required for MscS activity.
\newblock \emph{The EMBO journal} 18:1730--1737.

\bibitem[Wood(1999)]{wood1999}
Wood, J.~M., 1999.
\newblock Osmosensing by {{Bacteria}}: {{Signals}} and {{Membrane-Based
  Sensors}}.
\newblock \emph{Microbiology and Molecular Biology Reviews} 63:230--262.

\bibitem[Oglecka et~al.(2014)Oglecka, Rangamani, Liedberg, Kraut, and
  Parikh]{oglecka2014}
Oglecka, K., P.~Rangamani, B.~Liedberg, R.~S. Kraut, and A.~N. Parikh, 2014.
\newblock Oscillatory phase separation in giant lipid vesicles induced by
  transmembrane osmotic differentials.
\newblock \emph{eLife} 3:e03695.

\bibitem[Peterlin and Arrigler(2008)]{peterlin2008}
Peterlin, P., and V.~Arrigler, 2008.
\newblock Electroformation in a flow chamber with solution exchange as a means
  of preparation of flaccid giant vesicles.
\newblock \emph{Colloids and Surfaces B: Biointerfaces} 64:77--87.

\bibitem[Ho et~al.(2016)Ho, Rangamani, Liedberg, and Parikh]{ho2016}
Ho, J. C.~S., P.~Rangamani, B.~Liedberg, and A.~N. Parikh, 2016.
\newblock Mixing {{Water}}, {{Transducing Energy}}, and {{Shaping Membranes}}:
  {{Autonomously Self-Regulating Giant Vesicles}}.
\newblock \emph{Langmuir} .

\bibitem[Koslov and Markin(1984)]{koslov1984}
Koslov, M.~M., and V.~S. Markin, 1984.
\newblock A theory of osmotic lysis of lipid vesicles.
\newblock \emph{Journal of Theoretical Biology} 109:17--39.

\bibitem[Litster(1975)]{litster1975}
Litster, J.~D., 1975.
\newblock Stability of lipid bilayers and red blood cell membranes.
\newblock \emph{Physics Letters A} 53:193--194.

\bibitem[Sandre et~al.(1999)Sandre, Moreaux, and Brochard-Wyart]{sandre1999}
Sandre, O., L.~Moreaux, and F.~Brochard-Wyart, 1999.
\newblock Dynamics of transient pores in stretched vesicles.
\newblock \emph{Proceedings of the National Academy of Sciences}
  96:10591--10596.

\bibitem[Brochard-Wyart et~al.(2000)Brochard-Wyart, {de Gennes}, and
  Sandre]{brochard-wyart2000}
Brochard-Wyart, F., P.~G. {de Gennes}, and O.~Sandre, 2000.
\newblock Transient pores in stretched vesicles: role of leak-out.
\newblock \emph{Physica A: Statistical Mechanics and its Applications}
  278:32--51.

\bibitem[Idiart and Levin(2004)]{idiart2004}
Idiart, M.~A., and Y.~Levin, 2004.
\newblock Rupture of a liposomal vesicle.
\newblock \emph{Physical Review E} 69:061922.

\bibitem[Evans et~al.(2003)Evans, Heinrich, Ludwig, and Rawicz]{evans2003}
Evans, E., V.~Heinrich, F.~Ludwig, and W.~Rawicz, 2003.
\newblock Dynamic {{Tension Spectroscopy}} and {{Strength}} of
  {{Biomembranes}}.
\newblock \emph{Biophysical Journal} 85:2342--2350.

\bibitem[Ryham et~al.(2011)Ryham, Berezovik, and Cohen]{ryham2011}
Ryham, R., I.~Berezovik, and F.~S. Cohen, 2011.
\newblock Aqueous {{Viscosity Is}} the {{Primary Source}} of {{Friction}} in
  {{Lipidic Pore~Dynamics}}.
\newblock \emph{Biophysical Journal} 101:2929--2938.

\bibitem[Rand(2004)]{rand2004}
Rand, R.~P., 2004.
\newblock Probing the role of water in protein conformation and function.
\newblock \emph{Philosophical Transactions of the Royal Society of London B:
  Biological Sciences} 359:1277--1285.

\bibitem[Diz-Mu{\~n}oz et~al.(2013)Diz-Mu{\~n}oz, Fletcher, and
  Weiner]{diz-munoz2013}
Diz-Mu{\~n}oz, A., D.~A. Fletcher, and O.~D. Weiner, 2013.
\newblock Use the force: membrane tension as an organizer of cell shape and
  motility.
\newblock \emph{Trends in Cell Biology} 23:47--53.

\bibitem[Stroka et~al.(2014)Stroka, Jiang, Chen, Tong, Wirtz, Sun, and
  Konstantopoulos]{stroka2014}
Stroka, K.~M., H.~Jiang, S.-H. Chen, Z.~Tong, D.~Wirtz, S.~X. Sun, and
  K.~Konstantopoulos, 2014.
\newblock Water {{Permeation Drives Tumor Cell Migration}} in {{Confined
  Microenvironments}}.
\newblock \emph{Cell} 157:611--623.

\bibitem[Porta et~al.(2015)Porta, Ghilardi, Pasini, Laurson, Alava, Zapperi,
  and Amar]{porta2015}
Porta, C. A. M.~L., A.~Ghilardi, M.~Pasini, L.~Laurson, M.~J. Alava,
  S.~Zapperi, and M.~B. Amar, 2015.
\newblock Osmotic stress affects functional properties of human melanoma cell
  lines.
\newblock \emph{The European Physical Journal Plus} 130:1--15.

\bibitem[Angelova et~al.(1992)Angelova, Sol{\'e}au, M{\'e}l{\'e}ard, Faucon,
  and Bothorel]{angelova1992}
Angelova, M.~I., S.~Sol{\'e}au, P.~M{\'e}l{\'e}ard, F.~Faucon, and P.~Bothorel,
  1992.
\newblock Preparation of giant vesicles by external {{AC}} electric fields.
  {{Kinetics}} and applications.
\newblock \emph{In} C.~Helm, M.~L{\"o}sche, and H.~M{\"o}hwald, editors, Trends
  in {{Colloid}} and {{Interface Science VI}}, {Steinkopff}, number~89 in
  Progress in Colloid \& Polymer Science, 127--131.

\bibitem[Ting et~al.(2011)Ting, Appel{\"o}, and Wang]{ting2011}
Ting, C.~L., D.~Appel{\"o}, and Z.-G. Wang, 2011.
\newblock Minimum {{Energy Path}} to {{Membrane Pore Formation}} and
  {{Rupture}}.
\newblock \emph{Physical Review Letters} 106:168101.

\bibitem[Bicout and Kats(2012)]{bicout2012}
Bicout, D.~J., and E.~Kats, 2012.
\newblock Rupture of a biomembrane under dynamic surface tension.
\newblock \emph{Physical Review E} 85:031905.

\bibitem[Kubo(1966)]{kubo1966}
Kubo, R., 1966.
\newblock The fluctuation-dissipation theorem.
\newblock \emph{Reports on progress in physics} 29:255.

\bibitem[Aubin and Ryham(2016)]{aubin2016}
Aubin, C.~A., and R.~J. Ryham, 2016.
\newblock Stokes flow for a shrinking pore.
\newblock \emph{Journal of Fluid Mechanics} 788:228--245.

\bibitem[Happel and Brenner(1983)]{happel1983}
Happel, J., and H.~Brenner, 1983.
\newblock Low {{Reynolds Number Hydrodynamics}}: {{With Special Applications}}
  to {{Particulate Media}}.
\newblock {Springer Science \& Business Media}.

\bibitem[Brochard et~al.(1976)Brochard, De~Gennes, and Pfeuty]{brochard1976}
Brochard, F., P.~De~Gennes, and P.~Pfeuty, 1976.
\newblock Surface tension and deformations of membrane structures: relation to
  two-dimensional phase transitions.
\newblock \emph{Journal de Physique} 37:1099--1104.

\bibitem[Evans and Rawicz(1990)]{evans1990}
Evans, E., and W.~Rawicz, 1990.
\newblock Entropy-driven tension and bending elasticity in condensed-fluid
  membranes.
\newblock \emph{Physical Review Letters} 64:2094--2097.

\bibitem[Ertel et~al.(1993)Ertel, Marangoni, Marsh, Hallett, and
  Wood]{ertel1993}
Ertel, A., A.~G. Marangoni, J.~Marsh, F.~R. Hallett, and J.~M. Wood, 1993.
\newblock Mechanical properties of vesicles. {{I}}. {{Coordinated}} analysis of
  osmotic swelling and lysis.
\newblock \emph{Biophysical Journal} 64:426--434.

\bibitem[Hallett et~al.(1993)Hallett, Marsh, Nickel, and Wood]{hallett1993}
Hallett, F.~R., J.~Marsh, B.~G. Nickel, and J.~M. Wood, 1993.
\newblock Mechanical properties of vesicles. {{II}}. {{A}} model for osmotic
  swelling and lysis.
\newblock \emph{Biophysical Journal} 64:435--442.

\bibitem[Karatekin et~al.(2003)Karatekin, Sandre, Guitouni, Borghi, Puech, and
  Brochard-Wyart]{karatekin2003a}
Karatekin, E., O.~Sandre, H.~Guitouni, N.~Borghi, P.-H. Puech, and
  F.~Brochard-Wyart, 2003.
\newblock Cascades of {{Transient Pores}} in {{Giant Vesicles}}: {{Line
  Tension}} and {{Transport}}.
\newblock \emph{Biophysical Journal} 84:1734--1749.

\bibitem[Bloom et~al.(1991)Bloom, Evans, and Mouritsen]{bloom1991}
Bloom, M., E.~Evans, and O.~G. Mouritsen, 1991.
\newblock Physical properties of the fluid lipid-bilayer component of cell
  membranes: a perspective.
\newblock \emph{Quarterly reviews of biophysics} 24:293--397.

\bibitem[Evans and Smith(2011)]{evans2011}
Evans, E., and B.~A. Smith, 2011.
\newblock Kinetics of hole nucleation in biomembrane rupture.
\newblock \emph{New Journal of Physics} 13:095010.

\bibitem[Popescu and Popescu(2008)]{popescu2008}
Popescu, D., and A.~G. Popescu, 2008.
\newblock The working of a pulsatory liposome.
\newblock \emph{Journal of Theoretical Biology} 254:515--519.

\bibitem[J{\"u}licher et~al.(1997)J{\"u}licher, Ajdari, and
  Prost]{julicher1997}
J{\"u}licher, F., A.~Ajdari, and J.~Prost, 1997.
\newblock Modeling molecular motors.
\newblock \emph{Reviews of Modern Physics} 69:1269.

\bibitem[Oster(2002)]{oster2002}
Oster, G., 2002.
\newblock Brownian ratchets: {{Darwin}}'s motors.
\newblock \emph{Nature} 417:25--25.

\bibitem[Elowitz et~al.(2002)Elowitz, Levine, Siggia, and Swain]{elowitz2002}
Elowitz, M.~B., A.~J. Levine, E.~D. Siggia, and P.~S. Swain, 2002.
\newblock Stochastic gene expression in a single cell.
\newblock \emph{Science} 297:1183--1186.

\bibitem[Turlier et~al.(2016)Turlier, Fedosov, Audoly, Auth, Gov, Sykes,
  Joanny, Gompper, and Betz]{turlier2016}
Turlier, H., D.~Fedosov, B.~Audoly, T.~Auth, N.~Gov, C.~Sykes, J.-F. Joanny,
  G.~Gompper, and T.~Betz, 2016.
\newblock Equilibrium physics breakdown reveals the active nature of red blood
  cell flickering.
\newblock \emph{Nature Physics} .

\bibitem[Helfrich and Servuss(1984)]{helfrich1984}
Helfrich, W., and R.-M. Servuss, 1984.
\newblock Undulations, steric interaction and cohesion of fluid membranes.
\newblock \emph{Il Nuovo Cimento D} 3:137--151.

\bibitem[Alberts et~al.(2014)Alberts, Johnson, Lewis, Morgan, Raff, Roberts,
  and Walter]{alberts2014}
Alberts, B., A.~Johnson, J.~Lewis, D.~Morgan, M.~Raff, K.~Roberts, and
  P.~Walter, 2014.
\newblock Molecular {{Biology}} of the {{Cell}}.
\newblock {Garland Science}, 6 edition edition.

\bibitem[Rangamani et~al.(2014)Rangamani, Mandadap, and Oster]{rangamani2014}
Rangamani, P., K.~K. Mandadap, and G.~Oster, 2014.
\newblock Protein-induced membrane curvature alters local membrane tension.
\newblock \emph{Biophysical journal} 107:751--762.

\end{thebibliography}


\begin{thebibliography}{10}
\providecommand{\url}[1]{\texttt{#1}}
\providecommand{\urlprefix}{ }

\bibitem[Davies(2012)]{davies2012}
Davies, E.~R., 2012.
\newblock Computer and {{Machine Vision}}: {{Theory}}, {{Algorithms}},
  {{Practicalities}}.
\newblock {Academic Press}.

\bibitem[Schneider et~al.(2012)Schneider, Rasband, Eliceiri,
  et~al.]{schneider2012}
Schneider, C.~A., W.~S. Rasband, K.~W. Eliceiri, et~al., 2012.
\newblock NIH Image to ImageJ: 25 years of image analysis.
\newblock \emph{Nat methods} 9:671--675.

\bibitem[Peterlin et~al.(2012)Peterlin, Arrigler, Diamant, and
  Haleva]{peterlin2012}
Peterlin, P., V.~Arrigler, H.~Diamant, and E.~Haleva, 2012.
\newblock Permeability of {{Phospholipid Membrane}} for {{Small Polar Molecules
  Determined}} from {{Osmotic Swelling}} of {{Giant Phospholipid Vesicles}}.
\newblock \emph{In} A.~Igli{\v c}, editor, Advances in {{Planar Lipid
  Bilayers}} and {{Liposomes}}, {Academic Press}, volume~16, 301--335.

\bibitem[Fettiplace and Haydon(1980)]{fettiplace1980}
Fettiplace, R., and D.~A. Haydon, 1980.
\newblock Water permeability of lipid membranes.
\newblock \emph{Physiological Reviews} 60:510--550.

\bibitem[Deamer and Bramhall(1986)]{deamer1986}
Deamer, D.~W., and J.~Bramhall, 1986.
\newblock Special {{Issue}}: {{LiposomesPermeability}} of lipid bilayers to
  water and ionic solutes.
\newblock \emph{Chemistry and Physics of Lipids} 40:167--188.

\bibitem[Seifert(2008)]{seifert2008}
Seifert, U., 2008.
\newblock Stochastic thermodynamics: principles and perspectives.
\newblock \emph{The European Physical Journal B} 64:423--431.

\bibitem[Olbrich et~al.(2000)Olbrich, Rawicz, Needham, and Evans]{olbrich2000}
Olbrich, K., W.~Rawicz, D.~Needham, and E.~Evans, 2000.
\newblock Water permeability and mechanical strength of polyunsaturated lipid
  bilayers.
\newblock \emph{Biophysical Journal} 79:321--327.

\bibitem[Portet and Dimova(2010)]{portet2010}
Portet, T., and R.~Dimova, 2010.
\newblock A {{New Method}} for {{Measuring Edge Tensions}} and {{Stability}} of
  {{Lipid Bilayers}}: {{Effect}} of {{Membrane Composition}}.
\newblock \emph{Biophysical Journal} 99:3264--3273.

\bibitem[Hormel et~al.(2014)Hormel, Kurihara, Brennan, Wozniak, and
  Parthasarathy]{hormel2014}
Hormel, T.~T., S.~Q. Kurihara, M.~K. Brennan, M.~C. Wozniak, and
  R.~Parthasarathy, 2014.
\newblock Measuring {{Lipid Membrane Viscosity Using Rotational}} and
  {{Translational Probe Diffusion}}.
\newblock \emph{Physical Review Letters} 112:188101.

\bibitem[Linder et~al.(1976)Linder, Nassimbeni, Polson, and
  Rodgers]{linder1976}
Linder, P.~W., L.~R. Nassimbeni, A.~Polson, and A.~L. Rodgers, 1976.
\newblock The diffusion coefficient of sucrose in water. {{A}} physical
  chemistry experiment.
\newblock \emph{Journal of Chemical Education} 53:330.

\end{thebibliography}

\section*{\hf SUPPORTING CITATIONS}
References (42-51) appear in the Supporting Material.

\bibliographystylesupp{biophysj}

\clearpage

\pagebreak
\setcounter{page}{1}

\bigskip
{\hf {\bf \hf \Large SUPPORTING MATERIAL:
 Pulsatile lipid vesicles under osmotic stress}\\ 
 
 {\normalsize Morgan Chabanon, James C.S. Ho, Bo Liedberg, Atul N. Parikh, Padmini Rangamani}}
 \bigskip

\section*{\hf Table of Content}
 {\bf
\begin{tabular}{p{0.94\textwidth} p{0.5\textwidth}} 
Supporting Materials and Methods & 1\\
Model Development and Simulations & 2\\
Derivation of the analytical relations between cycle period, strain rate, and concentration differential & 4\\
Supporting References & 5\\
Supporting Figures and Movies & 6\\
\end{tabular} 
}

\section*{\hf Supporting Materials and Methods}

\textbf{Swell-burst cycle experiments. }
The experimental methods for the GUVs preparation has been described in \cite{oglecka2014, angelova1992}. Briefly, GUVs (100$\%$ POPC + 1mol$\%$ Rho-DPPE) containing 200 mM of sucrose were prepared by electroformation, yielding vesicles with radii ranging from 7 to 20~$\mu$m. GUVs were then placed in a bath of deionized water at room temperature, inducing hypotonic stress proportional to the inner sucrose concentration. The kinetics of eight GUVs were recorded by time-lapse microscopy at 1/150 images/ms. In order to allow for the sedimentation of GUVs to the bottom of the well, observations were started about one minute after the GUVs were subject to hypotonic conditions.

For each frame, the GUV radius was measured using a customized MATLAB (Mathworks, Natick, MA) code to streamline the image analysis. This code uses a Circular Hough Transform method based on a phase-coding algorithm to detect circles \citesupp{davies2012}, and measure their radii and centers. For our data, this custom code gives the evolution of the GUV radius in time with a precision of about 0.1 $\mu$m. Due to slow movement of the GUVs, in some cases the observation fields had to be adjusted to follow the GUVs, and the recording was paused. These are indicated by black dashed lines in Fig.~S\ref{figS_exp}. In order to define a systematic experimental initial GUV radius, $R_0$ was determined for each GUV as 0.995 times the first measured local minimum GUV radius, in accordance with our numerical results. Furthermore, burst events were identified by drops of GUV radius larger than 0.2 $\mu$m within a 1.5 s interval, and are plotted as solid red triangle. Bursting events that were likely to happen during the video gaps were indicated by plain red triangle (these ``likely" bursting events were not taken into account in the data processing for Figs.~\ref{fig5_1} and \ref{fig5_2}).

\textbf{Leak-out quantification. }
To quantify the leak-out amount when a membrane pore is formed, giant unilamellar vesicles (GUVs) were electroformed in 200 mM sucrose, supplemented with 58.4 $\mu$M 2-NBDG (2-(N- (7-Nitrobenz-2-oxa-1,3-diazol-4-yl) Amino) -2-Deoxyglucose), a fluorescent glucose analog that has an almost identical molecular weight as sucrose. Fluorescence imaging was performed on a deconvolution microscope, equipped with a FITC filter. Time-lapse imaging of the vesicles was performed approximately one minute after exposing the vesicles to either deionized water (hypotonic conditions $n=3$) or glucose (isotonic conditions $n=3$) environment to ensure sedimentation of GUVs to the bottom of the well. All acquisitions were performed using identical settings to facilitate comparison of vesicles submerged in water or equi-osmotic glucose environment. 

For Fig.~\ref{fig4}(C), the GUVs were detected with a MATLAB (Mathworks, Natick, MA) code adapted from the one described above, where the mean gray intensity inside and outside of the GUV are measured. For every time frame, the difference between the inner and outer mean intensity $\Delta I(t)$ was computed, and normalized by the intensity difference of the first frame  $\Delta I(t_0)$. Bursting events were identified by visual inspection of the videos, and reported by arrows on Fig.~\ref{fig4}(B).

In order to highlight the efflux of fluorescent dyes during a GUV bursting event, three frames (before, during and after the event) were extracted form the video of a GUV containing 200~mM sucrose + 58.4~$\mu$M 2-NBDG in hypotonic conditions (Fig.~\ref{fig4}(C)). These images were further processed with ImageJ software \citesupp{schneider2012} to plot Fig.~\ref{fig4}(D). Briefly, the noise was attenuated by successively applying ImageJ built-in routines (background suppression, contrast enhancing, median filter), and ploting the isovalues of gray at 90, 75 and 60 with the pluggin Contour Plotter (http://rsb.info.nih.gov/ij/plugins/contour-plotter.html).

\textbf{Burst cycle analysis. }
\revise{
The following analysis has been applied for both experimental and numerical data in order to produce Figs.~\ref{fig5_1}, \ref{fig5_2} and S\ref{figS_comparison}.
For a given GUV radius dynamics, a swell-burst cycle was defined between two successive minimum GUV radii that immediately followed a bursting event. Cycle periods were computed as the time between two consecutive minima in vesicular radii. For experimental data, if there was a video gap between two consecutive radius minima, the cycle was not taken into account for the analysis, that is to say, only cycles between two successive solid triangles in Figs.~\ref{fig1}(C) and S\ref{figS_exp} were taken into account. The strain rate was computed as the difference between the maximum and minimum radii within these cycles, divided by the time between these two events. For experimental data, because of the lag between the beginning of the experiments and the beginning of the video recordings, the initial observed cycle number was adjusted between $n=1$ and $n=4$, depending on $R_0$.}

\section*{\hf Model Development and Simulations}

Here we derive a theoretical model to describe the swell-burst cycle of a GUV under hypotonic conditions. In line with previous work \cite{koslov1984, idiart2004, popescu2008}
, the model has three conservation equations, governing the dynamics of the solvent, solute, and membrane pore.

\textbf{Mass conservation of solvent. }
Mass conservation of the solvent (water) within the vesicle is governed by the flux through the membrane ($j_w$), and the leak-out through the pore. For a spherical GUV, the general form of the mass conservation equation for the solvent is
\begin{equation} \label{eq:sol_mass_cons1}
\frac{d}{dt}\left( \frac{4}{3}\pi R^3 \rho_s \right) = j_w - \pi r^2 \rho_s v_L \;,
\end{equation}
where $R$ and $r$ are the radius of the vesicle and the pore respectively, $\rho_s$ is the mass density of the solvent, and $v_L$ is the leak-out velocity of the solvent. The osmotic flux is influenced by the permeability of the membrane to the solvent ($P$), the osmotic pressure ($\Delta p_{osm}$), and the Laplace pressure ($\Delta p_L$). A phenomenological expression for the osmotic flux is \cite{koslov1984, popescu2008}\citesupp{peterlin2012}
\begin{equation} \label{eq:jw}
j_w = \frac{P \nu_s \rho_s}{k_B T N_A} A \left( \Delta p_{osm} - \Delta p_L \right) \;,
\end{equation}
where $\nu_s$ is the solvent molar volume, and the membrane area is defined as $A = 4\pi R^2 - \pi r^2$. The two pressures involved in Eq.~\ref{eq:jw} are defined as
\begin{equation}
\left\{
\begin{array}{l}
\Delta p_{osm} = k_B T N_A \Delta c \\
\Delta p_L = \dfrac{2\sigma}{R}
\end{array}\right. \;.
\end{equation}
The Laplace pressure originates from the surface tension in the membrane $\sigma$, which we assume to be proportional to the membrane strain
\begin{equation} \label{eq:sigma_linear}
\sigma = \kappa_\text{eff} \frac{A-A_0}{A_0} = \kappa_\text{eff}  \epsilon \;.
\end{equation}
Here $\kappa_\text{eff}$ is the effective area extension modulus (combining the effects of membrane unfolding and elastic deformation), and $A_0 = 4\pi R_0^2$ is the surface of the vesicle in its unstretched state.
The leak-out velocity $v_L$ can be analytically approximated at low Reynolds number in order to relate it to the Laplace pressure \cite{happel1983}
\begin{equation} \label{eq:vl}
v_L = \frac{\Delta p_L r}{3\pi\eta_s} \;.
\end{equation}

Substituting these definitions into Eq.~\ref{eq:sol_mass_cons1}, the mass conservation equation for the solvent takes the form of an ordinary differential equation (ODE) for the GUV radius
\begin{equation} \label{eq:sol_mass_cons2}
4\pi R^2\frac{dR}{dt}= \frac{P \nu_s }{k_B T N_A} A \left( k_B T N_A \Delta c - \frac{2\sigma}{R}  \right) -   \frac{2\sigma }{3\eta_s R} r^3 \;.
\end{equation}

\textbf{Mass conservation of solute. }
The permeability of lipid membranes to water is several orders of magnitude larger than for most solutes \citesupp{fettiplace1980, deamer1986}. Consequently the lipid bilayer is supposed to be semi-permeable, neglecting sucrose transport through the membrane. Thus, variation of solute in the vesicle is exclusively limited to diffusive and convective transport through the pore, such that
\begin{equation} \label{eq:solute_mass_cons1}
\frac{d}{dt}\left(\frac{4}{3}\pi R^3 \Delta c \right) = \pi r^2 \left( - D \frac{\Delta c}{R} - v_L\Delta c \right) \;.
\end{equation}
While the diffusive flux through the pore is usually neglected over the convective efflux of solute, theoretical analysis of long lived pores indicates that the Laplace pressure decreases rapidly after the pore opening, and stays low for most of the pore life time \cite{brochard-wyart2000}. This suggests that the convective efflux directed by the leak-out velocity may not always be the dominant solute transport mechanism, as confirmed by our numerical and experimental results (see  main text Fig.~\ref{fig4}). Expanding Eq.~\ref{eq:solute_mass_cons1} we obtain an ODE for the concentration difference in solute
\begin{equation} \label{eq:solute_mass_cons2}
\frac{4}{3}\pi R^3 \frac{d \Delta c }{dt} = - D\pi r^2  \frac{\Delta c}{R} - \frac{2\sigma }{3\eta_s R} r^3\Delta c -  4\pi R^2\Delta c \frac{d R}{dt} \;.
\end{equation}

\textbf{Pore force balance. }
The pore in the lipid bilayer is modeled as an overdamped system, where the pore radius is governed by the following Langevin equation \citesupp{seifert2008}
\begin{equation}\label{eq:pore_energy1}
\zeta \frac{d}{dt}\left( 2\pi r \right) = F(r,t) + \xi(t) \;,
\end{equation}
where $\zeta$ is the membrane drag coefficient (inverse of the mobility), $F(r,t)$ is a conservative force, and $\xi$ is a noise term accounting for independent thermally-induced pore fluctuations. The drag coefficient includes two in-plane contributions $\zeta = \zeta_m + \zeta_s$: one from membrane dissipation, proportional to the membrane viscosity and thickness $\zeta_m=\eta_m h$ \cite{brochard-wyart2000}, and a second from the friction of the solvent with the moving pore -- proportional  to the solvent viscosity $\zeta_s = C \eta_s r $, where $C=2\pi$ is a geometric coefficient \cite{ryham2011, aubin2016}.
The conservative force $F(r,t)=-\partial V(r,t)/\partial r$ arises from the membrane potential $V(r,t)$, which is equal to the sum of the strain energy $V_s$, and the pore energy $V_p$. We assume the membrane strain energy to take a Hookean form $V_s = \kappa_\text{eff} \left(A-A_0\right)^2 /(2 A_0)$, where $\kappa_\text{eff}$ is an effective stretching modulus approximating the combined contributions of membrane unfolding and elastic stretching. The pore energy depends on the edge energy and length as $V_p = 2\pi r \gamma$,
where $\gamma$ is the pore line tension, here assumed independent of the pore radius. Using the definition $\sigma = \partial V_s/\partial A$, we can therefore express the force as
\begin{equation}
F(r,t) =2\pi\sigma r - 2\pi\gamma \;.
\end{equation}
The fluctuation term has a zero mean, and a correlation function given by
\begin{equation}
\langle \xi(t)\xi(t^\prime) \rangle = 2\zeta k_B T \delta(t-t^\prime) \;,
\end{equation}
following the dissipation-fluctuation theorem, where $\delta$ is the Dirac delta function.

Rearranging Eq.~\ref{eq:pore_energy1} with these definitions, we obtain a stochastic differential equation for the pore radius
\begin{equation}\label{eq:pore_energy2}
\left(\eta_m h + C \eta_s r\right)\frac{d}{dt}(2\pi r) = 2\pi(\sigma r - \gamma) + \xi(t) \;,
\end{equation}
with $r\ge0$. The last term in Eq.~\ref{eq:pore_energy2} is responsible for thermally driven pore nucleation. 

\textbf{Deterministic model. }
In the absence of thermal fluctuations, a critical value for the membrane tension (or strain) has to be defined, and an initial pore has to be set artificially in order for a large pore to open. In that case, Eq.~\ref{eq:pore_energy2} is simply
\begin{equation}\label{eq:pore_energy_det}
\left(\eta_m h + C \eta_s r\right)\frac{dr}{dt} = \sigma r - \gamma \;.
\end{equation}
When the pore is closed ($r=0$) and the strain overcomes the predetermined critical value ($\epsilon\ge\epsilon^*$), an initial pore large enough to overcome the nucleation barrier ($r=\gamma/\sigma$) is artificially created.

\textbf{Numerical implementation. }
All numerical computations have been carried out using a custom code in MATLAB (Mathworks, Natick, MA). The stochastic model, composed of Eqs.~\ref{eq:sde_r}, \ref{eq:ode_R} and \ref{eq:ode_c} was solved using an order-1 Runge-Kutta scheme. Because a pore nucleation event occurs due to a single fluctuation overcoming the energy barrier, the numerical implementation of the noise requires the definition of a fluctuation frequency $f_T$ \revise{(number of fluctuation "kicks" per seconds)} that is independent of the time step. For comparison a deterministic model (Eqs.~\ref{eq:sol_mass_cons2}, \ref{eq:solute_mass_cons2}, and \ref{eq:pore_energy_det}) was solved using Euler method. All parameters are shown in Table~S\ref{tab_parameters}. All time steps were taken as 0.1 ms, (smaller time steps did not improve the accuracy of the results significantly). For the cycle analysis of the stochastic model, Figs.~\ref{fig5_1} and \ref{fig5_2}, shows the average and standard deviations of 10 runs with same parameters.

\begin{supptable}[tbp]
\centering
\begin{tabular}{l l l}
Parameter & Typical value & References \\
\hline
$R_0$ & 8-20 $\mu$m & this work\\ 
$c_0$ & 200 mM & this work \\ 
$d$   & 3.5 nm  & \\ 
$\rho_s$  & 1000  kg m$^{-3}$ &  \\ 
$\nu_s$  & 18.04$\times$10$^{-6}$  m$^3$ mol$^{-1}$ &  \\ 
$P$ & 20 $\mu$m/s  & \citesupp{olbrich2000} \\
$T$ & 294 K  & \\
$\gamma$  & 5 pN  & \citesupp{portet2010} \\
$\kappa_\text{eff}$  & 2$\times$10$^{-3}$ N/m  &  this work \\
$\eta_m$  & 5 Pa s  & \citesupp{hormel2014}  \\ 
$\eta_s$  & 0.001 Pa s &  \\ 
$D$  &  5$\times$10$^{-10}$ m$^2$/s & \citesupp{linder1976} \\ 
$C$ & $2\pi$ & \cite{aubin2016} \\ 
$f_T$ & 150 Hz & this work\\
\hline
\end{tabular} 
\caption{Parameters used in the simulations} \label{tab_parameters} 
\end{supptable}

\section*{\hf Derivation of the analytical relations between cycle period, strain rate, and concentration differential}

First, we derive the linear dependence of the strain rate on the concentration difference shown in Fig.~\ref{fig5_2}(D). For a closed vesicle ($r=0$), the membrane area is $A=4\pi R^2$, and the strain rate is 
\begin{equation}
\dot{\epsilon} = \frac{d}{dt} \left( \frac{A - A_0}{A_0} \right) = \frac{2R}{R_0}\frac{dR}{dt} \;.
\end{equation}
This allows us to write Eq.~\ref{eq:sol_mass_cons2} in terms of the strain rate as
\begin{equation} 
\dot{\epsilon} = \frac{P \nu_s }{k_B T N_A} \frac{A}{2\pi R R_0^2} \left( k_B T N_A \Delta c - \frac{2\sigma}{R}  \right)\;.
\end{equation}
When the osmotic pressure is the dominant process influencing GUV swelling, we can neglect the Laplace pressure and obtain
\begin{equation} \label{eq:eps_dot}
\tau \dot{\epsilon} \simeq \frac{2R}{R_0}\frac{\Delta c}{c_0} \;,
\end{equation}
where $\tau=R_0/(P \nu_s c_0)$. At maximum GUV radius amplitude, $R/R_0$ can be expressed in term of the lytic strain as $R_\text{max}/R_0 = \sqrt{\epsilon^*+1}$, allowing to write Eq.~\ref{eq:eps_dot} as
\begin{equation} \label{eq:eps_dot_s}
\tau \dot{\epsilon} \simeq 2\sqrt{\epsilon^*+1}\frac{\Delta c}{c_0} \;.
\end{equation}
Plotting this relationship in Figs.~\ref{fig5_1}(D) and \ref{fig5_2}(D) for a typical lytic strain $\epsilon^*=0.15$, we get a good agreement with the numerical results from the stochastic model.

We now derive an approximate relation between the cycle period and the strain rate. During a cycle of period $T_n$, the lytic strain can be written
\begin{equation}
\epsilon^* \simeq T_n \dot{\epsilon} \;.
\end{equation}
Introducing Eq.~\ref{eq:eps_dot_s}, we get
\begin{equation}
\frac{T_n}{\tau} \simeq \frac{\epsilon^*}{2\sqrt{\epsilon^*+1}} \left( \frac{\Delta c}{c_0}\right)^{-1} \;.
\end{equation}
Taking $\epsilon^*=0.15$, this relationship fits well the simulation results, as shown in Figs.~\ref{fig5_1}(C) and \ref{fig5_2}(C).

It should be noted that, because the Laplace pressure is neglected in the derivation of Eq.~\ref{eq:eps_dot_s}, the analytical expression slightly overestimates the strain rate as shown in Figs.~\ref{fig5_1}(D) and \ref{fig5_2}(D). Moreover the cycle period is also overestimated for low solute concentrations due to the constant lytic strain assumed in the analytical expression (Figs.~\ref{fig5_1}(C) and \ref{fig5_2}(C)).


\bibliographystylesupp{biophysj}
\bibliographysupp{biblio.bib}

\section*{\hf Supporting Figures and Movies}


\paragraph{Supporting Movie 1} Membrane nodules appearance after membrane pore reseals. Movie assembled from time-lapse fluorescence microscopy images (frame rate, 2 fps; total duration, 17 s; image size, 82.43 $\mu$m $\times$ 82.43 $\mu$m; scale bar, 10 $\mu$m) obtained for a population of electroformed GUVs consisting of POPC doped with 1$\%$ Rhodamine-B labeled DPPE membrane in a hypotonic solution (Osmotic differential of 200 mM).   

\paragraph{Supporting Movie 2} Multiple swell-burst cycles of GUVs subject to hypotonic stress. Movie assembled from time-lapse fluorescence microscopy images (frame rate, 24 fps; total duration, 77 s; image size, 82.43 $\mu$m $\times$ 82.43 $\mu$m; scale bar, 10 $\mu$m) obtained for a population of electroformed GUVs consisting of POPC doped with 1$\%$ Rhodamine-B labeled DPPE membrane in a hypotonic solution (Osmotic differential of 200 mM). 

\paragraph{Supporting Movie 3} Model results showing multiple swell-burst cycles of a GUV subject to hypotonic stress.  GUV radius (top-left panel), pore radius (middle-left panel), and solute differential (bottom-left panel) as a function of time. Right panel is a representation of the numerical GUV in time, where the grey intensity is proportional to the inner sucrose concentration.
GUV initial radius is $R_0=14\;\mu$m, initial solute concentration is $c_0=200$~mM. All parameters are shown in Supporting Table~S\ref{tab_parameters}.

\paragraph{Supporting Movie 4} Model results showing a single pore opening dynamics of a GUV subject to hypotonic stress.  GUV radius (top-left panel), pore radius (middle-left panel), and solute differential (bottom-left panel) as a function of time. Right panel is a representation of the numerical GUV in time, where the grey intensity is proportional to the inner sucrose concentration.
GUV initial radius is $R_0=14\;\mu$m, initial solute concentration is $c_0=200$~mM. All parameters are shown in Supporting Table~\ref{tab_parameters}.

\paragraph{Supporting Movie 5} Solute leakage of a GUV in multiple swell-burst cycles under hypotonic condition. Movie assembled from time-lapse fluorescence microscopy images (frame rate, 24 fps; total duration, 11 s; image size, 119.14 $\mu$m $\times$ 125.58 $\mu$m; scale bar, 20 $\mu$m) obtained for a population of electroformed GUVs consisting of POPC doped with 1$\%$ Rhodamine-B labeled DPPE membrane in a hypotonic solution (Osmotic differential of 200 mM). 

\paragraph{Supporting Movie 6} GUV under isotonic condition. Movie assembled from time-lapse fluorescence microscopy images (frame rate, 24 fps; total duration, 8 s; image size, 101.11 $\mu$m $\times$ 101.11 $\mu$m; scale bar, 10 $\mu$m) obtained for a population of electroformed GUVs consisting of POPC doped with 1$\%$ Rhodamine-B labeled DPPE membrane in a isotonic solution (no osmotic differential).  

\paragraph{Supporting Movie 7} Solute efflux from GUV during one swell-burst cycle. Movie assembled from time-lapse fluorescence microscopy images (frame rate, 12 fps; total duration, 8 s; image size, 164.86 $\mu$m $\times$ 164.86 $\mu$m; scale bar, 20 $\mu$m) obtained for a population of electroformed GUVs consisting of POPC doped with 1$\%$ Rhodamine-B labeled DPPE membrane in a hypotonic solution (Osmotic differential of 200 mM).

%


\begin{suppfigure*}[htbp] 
\centering
\includegraphics[width=1\textwidth]{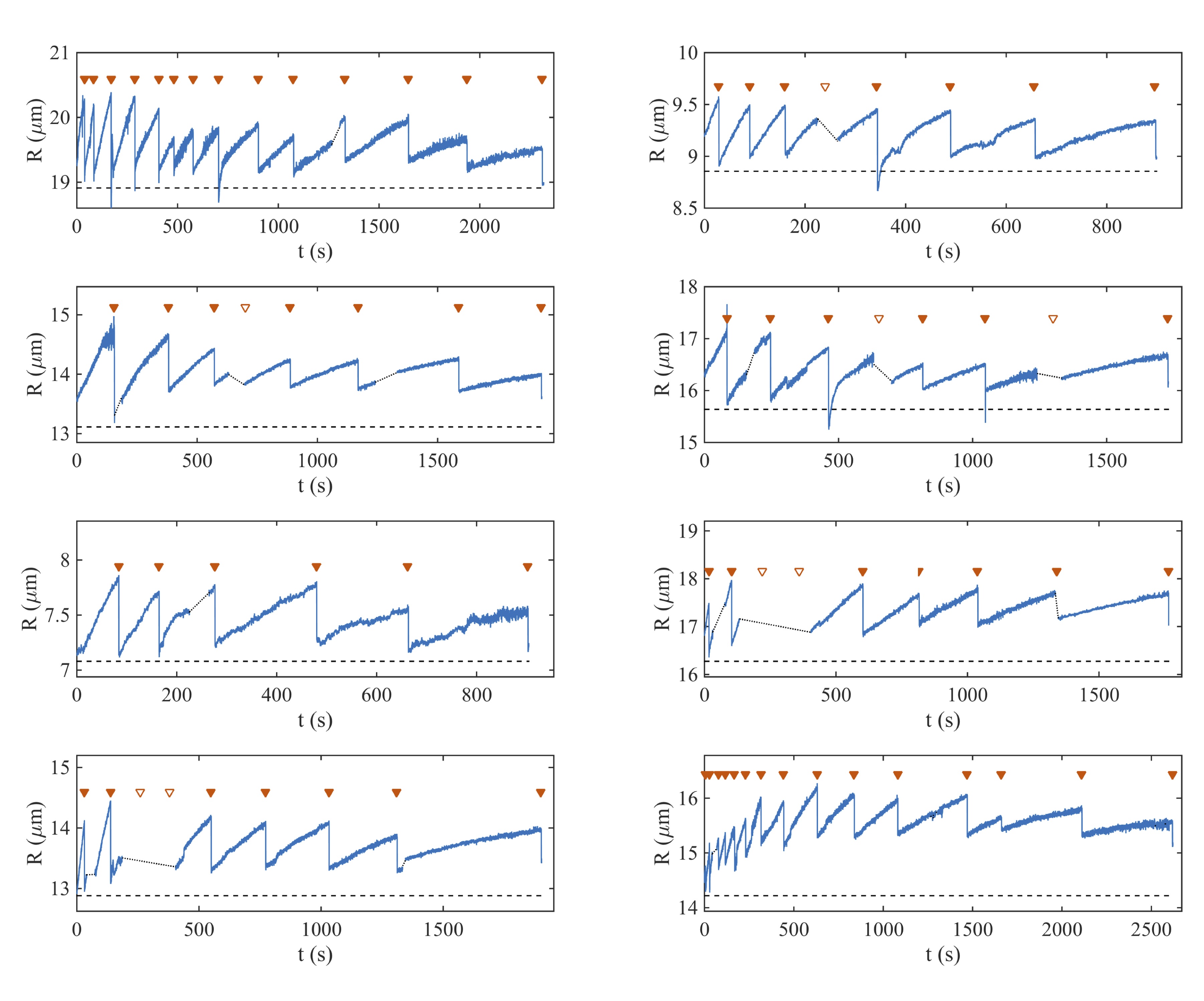}
\caption{Experimental measurements of GUV radius during swell-burst cycles in 200~mM sucrose hypotonic solutions. Videos of GUV were recorded and analyzed with a contour detection software, as described in the Material and Methods section. The radii of eight GUVs from different experiments are plotted here as a function of time. The radii increase continuously during swelling phases, and drop abruptly when bursting events occur. Each observed pore opening event is indicated by a red triangle. Gaps in the videos due to experimental constraints are shown by dashed lines, and pore opening events that likely occurred during these gaps are indicated by red triangle.}
\label{figS_exp}
\end{suppfigure*}

\begin{suppfigure*}[htbp] 
\centering
\includegraphics[width=0.5\textwidth]{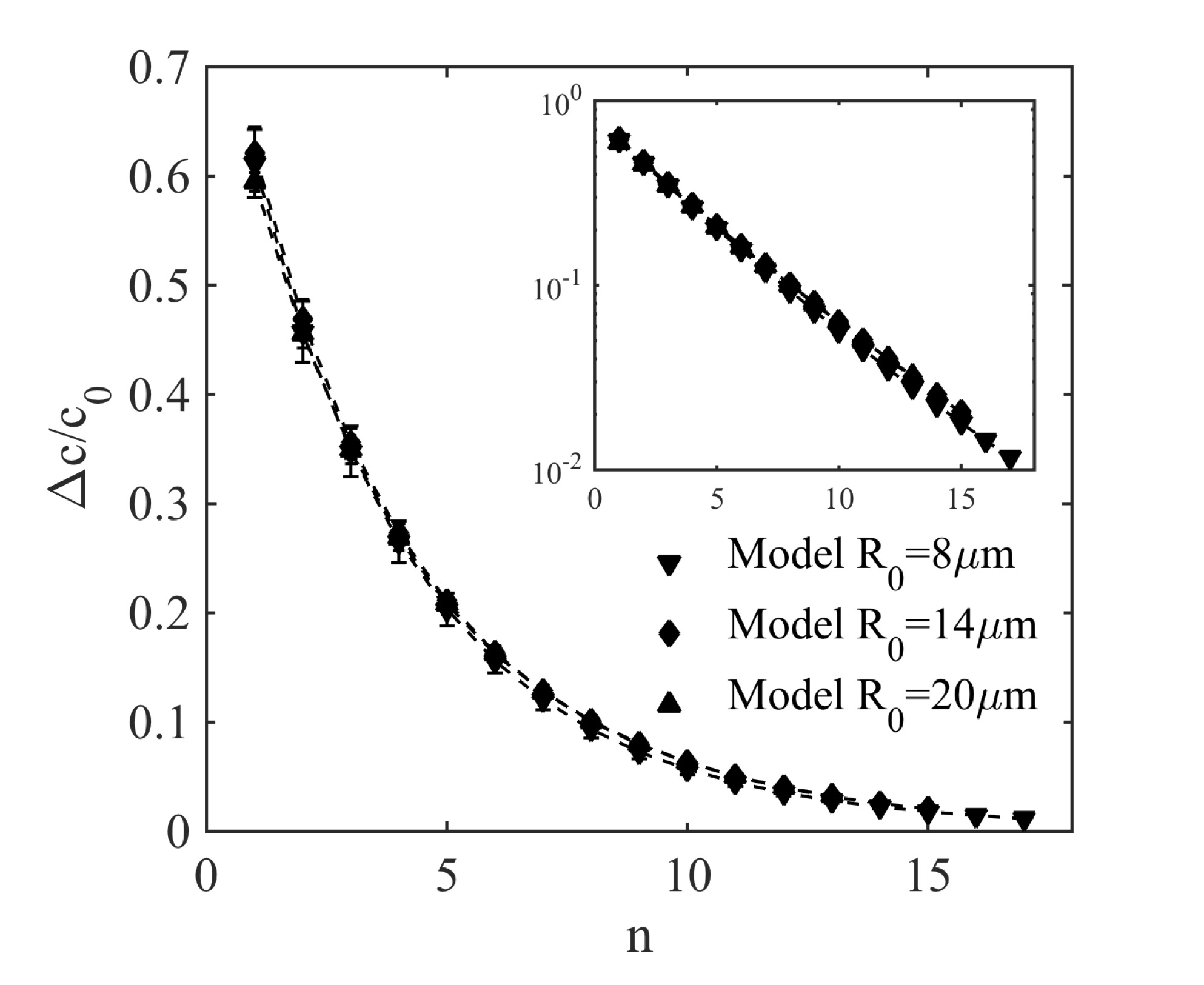}
\caption{Model results for the solute concentration as a function of the cycle number. Insets shows the same data with a logarithmic y axis, highlighting the exponential dependence $\ln(\Delta c/c_0) \simeq-0.25n-0.33$.}
\label{figS_cn}
\end{suppfigure*}

\begin{suppfigure*}[htbp] 
\centering
\includegraphics[width=1\textwidth]{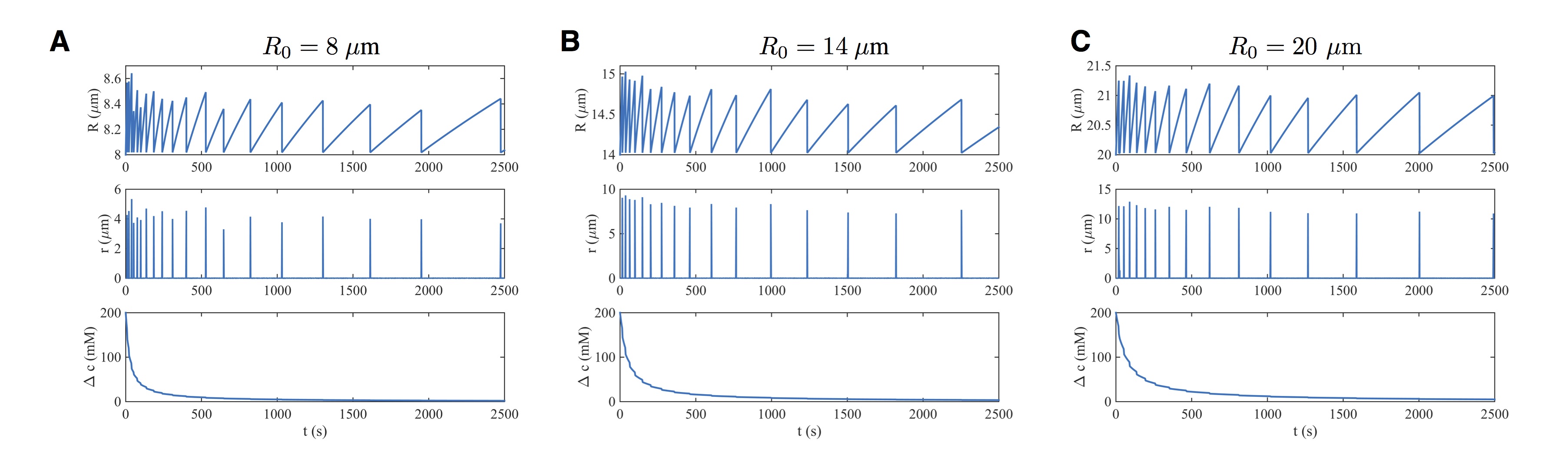}
\caption{Influence of resting radius on the GUV swell-burst dynamics. All parameters are as shown in Table~\ref{tab_parameters}, except in panels (A and C) where $R_0$ is set as indicated.}
\label{figS_R}
\end{suppfigure*}

\begin{suppfigure*}[htbp] 
\centering
\includegraphics[width=1\textwidth]{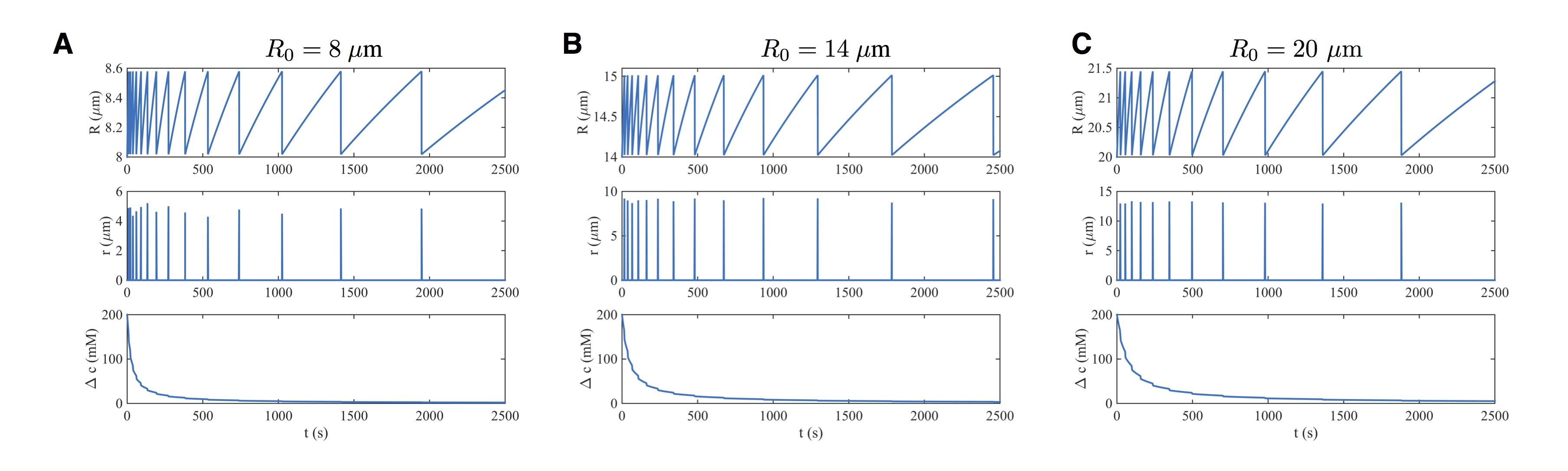}
\caption{Solution to the deterministic model. All parameters are as shown in Table~\ref{tab_parameters} except the pore opening strain set to $15\%$.}
\label{figS_det}
\end{suppfigure*}

\begin{suppfigure*}[htbp] 
\centering
\includegraphics[width=1\textwidth]{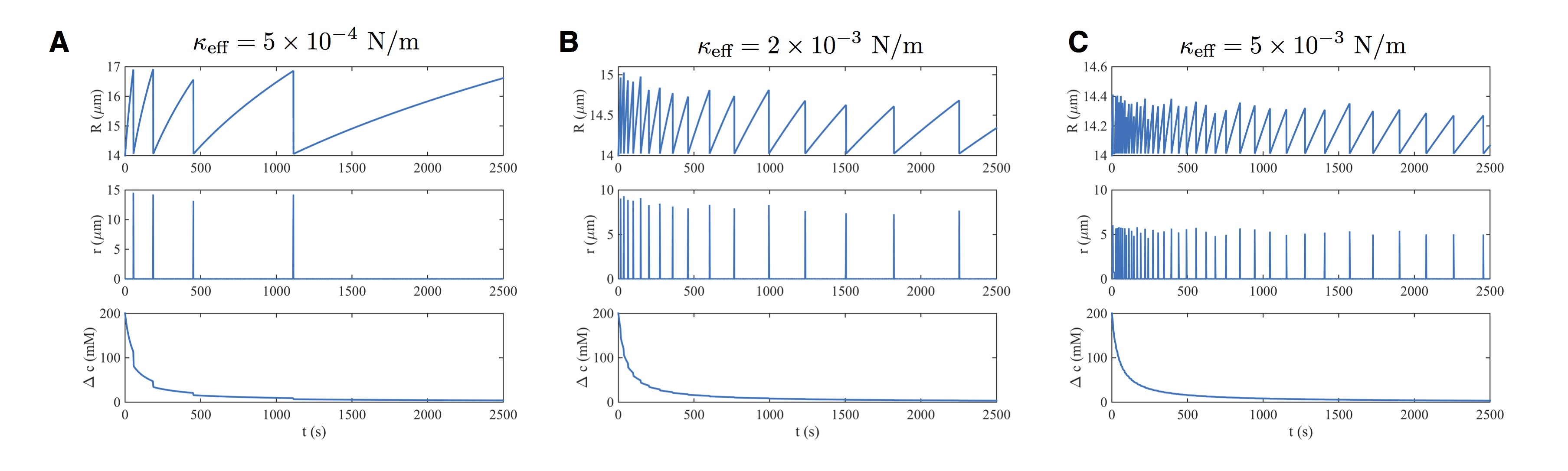}
\caption{Influence of the effective stretching modulus on the GUV swell-burst dynamics. All parameters are as shown in Table~\ref{tab_parameters}, except in panels (A and C) where $\kappa_\text{eff}$ is set as indicated, and $R_0=14\;\mu m$.}
\label{figS_kappa}
\end{suppfigure*}

\begin{suppfigure*}[htbp] 
\centering
\includegraphics[width=1\textwidth]{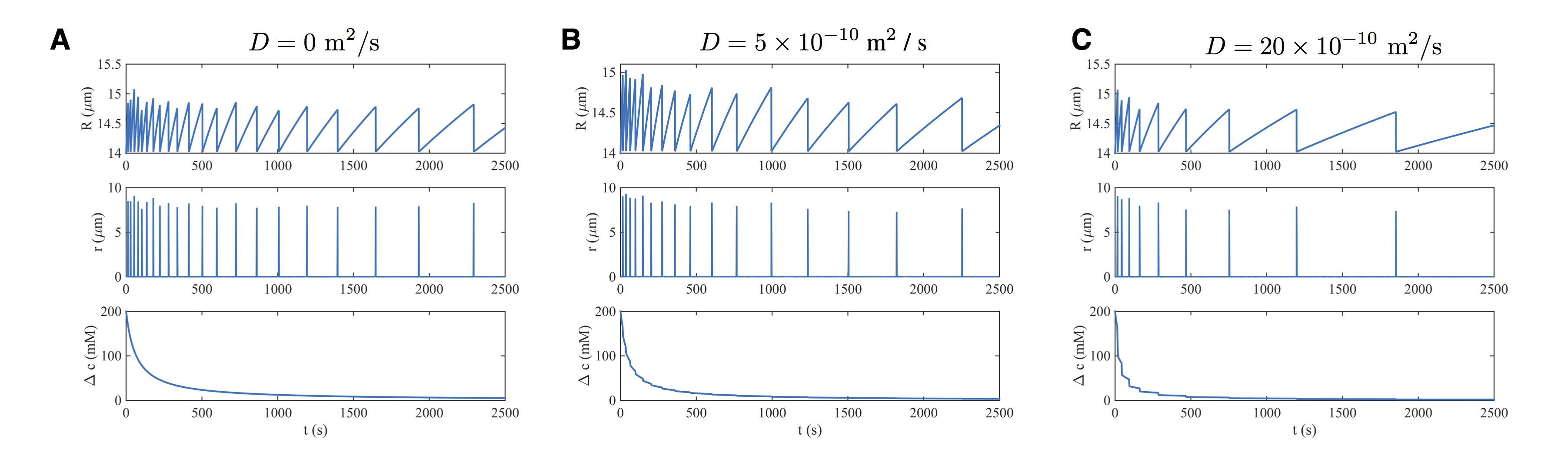}
\caption{Influence of solute diffusion on the GUV swell-burst dynamics.  All parameters are as shown in Table~\ref{tab_parameters}, except in panels (A and C) where $D$ is set as indicated, and $R_0=14\;\mu m$.}
\label{figS_D}
\end{suppfigure*}

\begin{suppfigure*}[htbp] 
\centering
\includegraphics[width=1\textwidth]{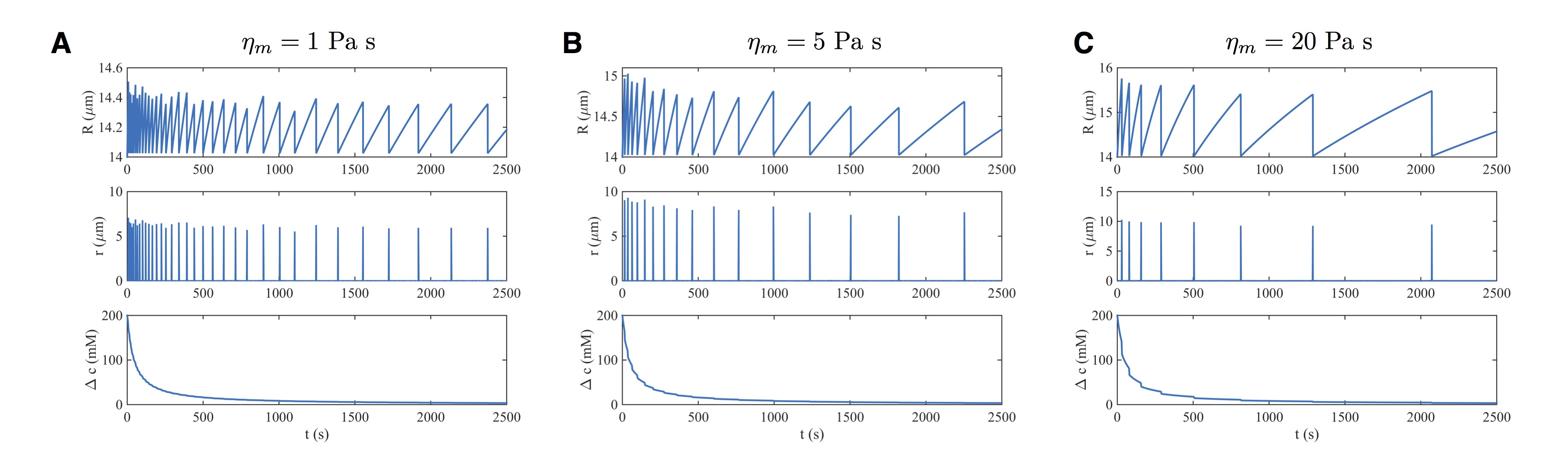}
\caption{Influence of membrane viscosity on the GUV swell-burst dynamics. All parameters are as shown in Table~\ref{tab_parameters}, except in panels (A and C) where $\eta_m$ is set as indicated, and $R_0=14\;\mu m$.}
\label{figS_etal}
\end{suppfigure*}

\begin{suppfigure*}[htbp] 
\centering
\includegraphics[width=1\textwidth]{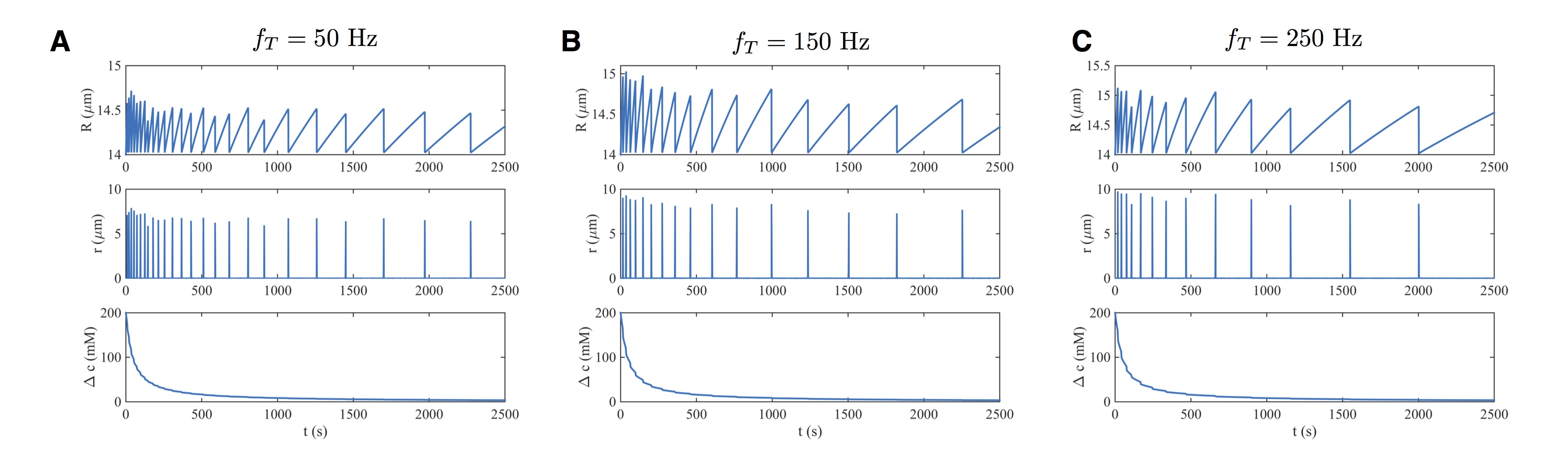}
\caption{Influence of fluctuation frequency on the GUV swell-burst dynamics. All parameters are as shown in Table~\ref{tab_parameters}, except in panels (A and C) where $f_T$ is set as indicated, and $R_0=14\;\mu m$.}
\label{figS_ft}
\end{suppfigure*}

\begin{suppfigure*}[htbp] 
\centering
\includegraphics[width=1\textwidth]{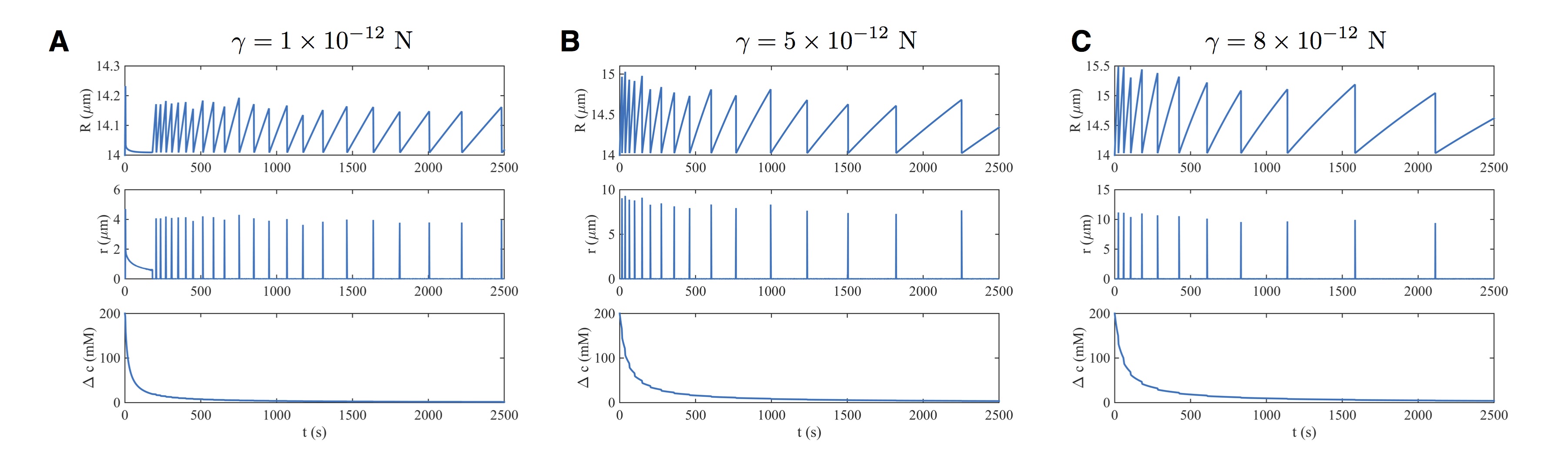}
\caption{Influence of membrane line tension on the GUV swell-burst dynamics. All parameters are as shown in Table~\ref{tab_parameters}, except in panels (A and C) where $\gamma$ is set as indicated, and $R_0=14\;\mu m$.}
\label{figS_gamma}
\end{suppfigure*}

\begin{suppfigure*}[htbp] 
\centering
\includegraphics[width=1\textwidth]{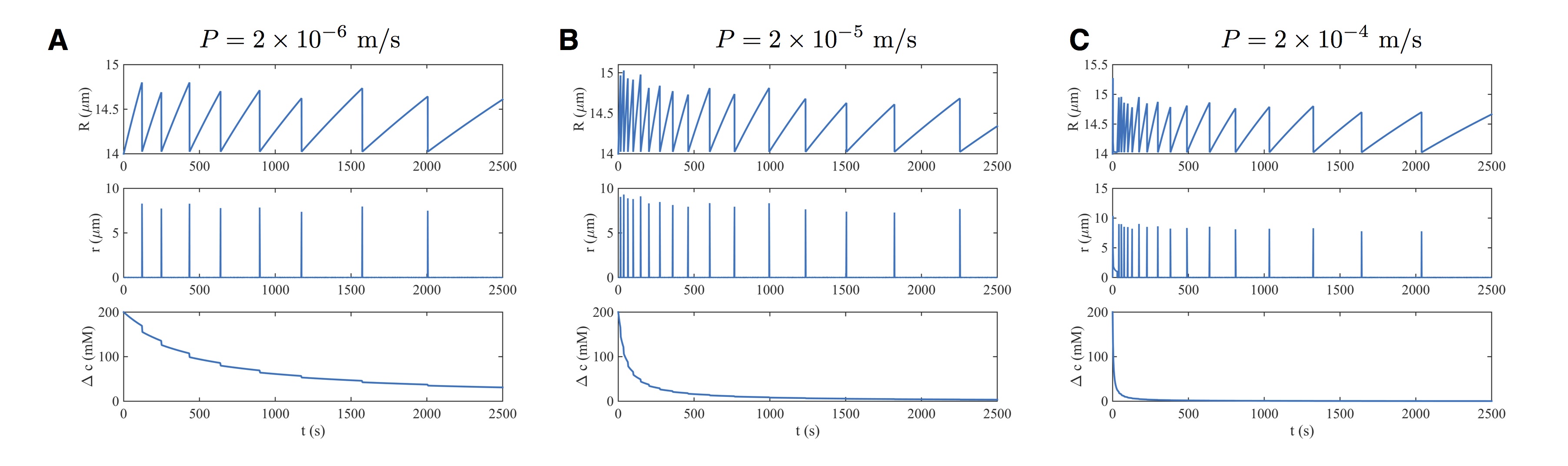}
\caption{Influence of membrane solvent permeability on the GUV swell-burst dynamics. All parameters are as shown in Table~\ref{tab_parameters}, except in panels (A and C) where $P$ is set as indicated, and $R_0=14\;\mu m$.}
\label{figS_P}
\end{suppfigure*}

\begin{suppfigure*}[htbp] 
\centering
\includegraphics[width=0.6\textwidth]{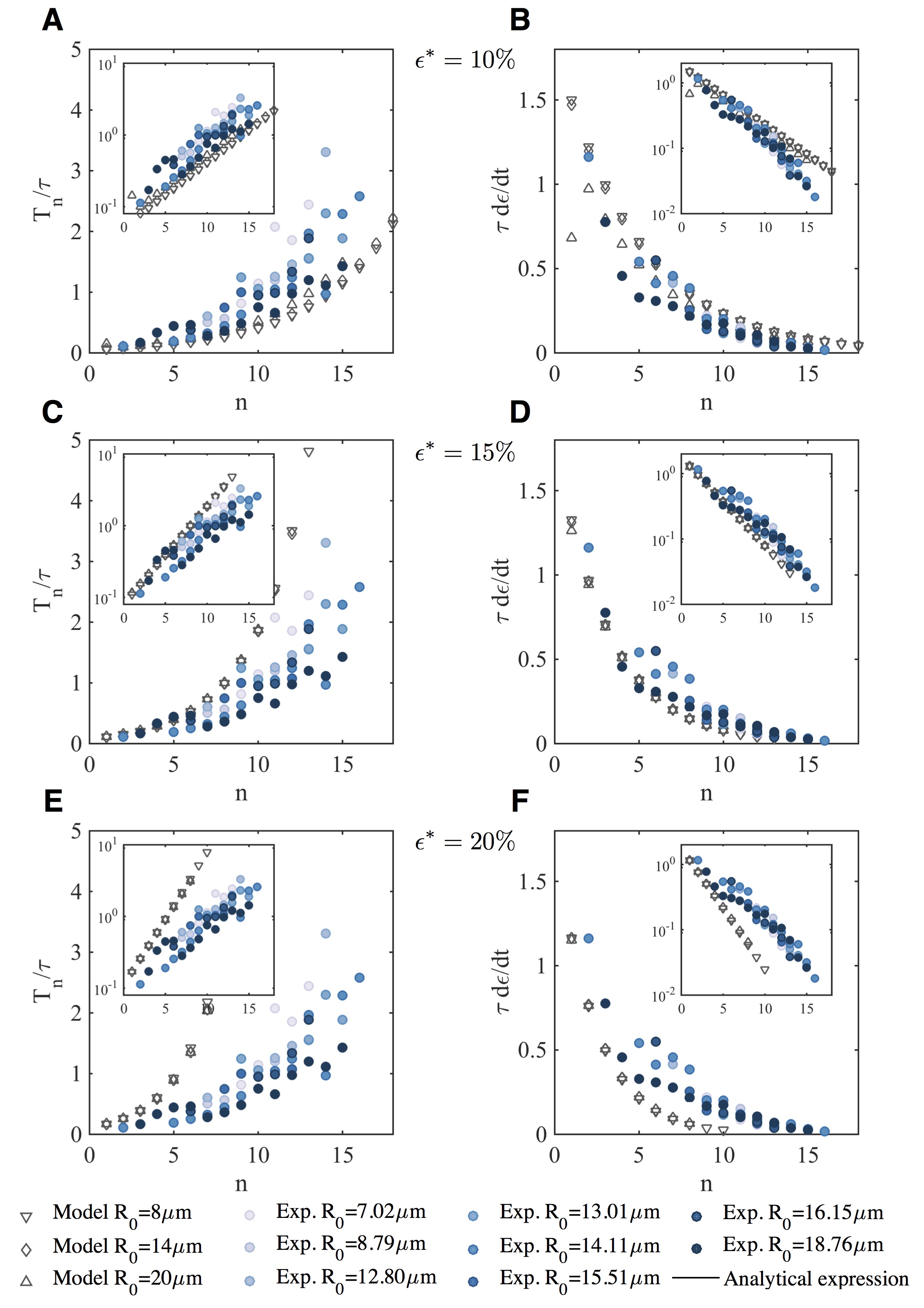}
\caption{The deterministic model (constant strain to rupture $\epsilon^*$) fails to match the experimental cycle period and strain rate. (A and B) $\epsilon^*=10\%$, (C and D) $\epsilon^*=15\%$, (E and F) $\epsilon^*=20\%$.}
\label{figS_comparison}
\end{suppfigure*}

\end{document}